\documentclass[a4paper,11pt]{article}
\pdfoutput=1
\usepackage{jcappub}

\usepackage{amsmath}
\usepackage{bm}
\usepackage{xcolor,color,colortbl}
\usepackage[utf8]{inputenc}
\usepackage{hyperref}
\usepackage{url}
\usepackage{graphicx}
\usepackage{multirow,bigdelim}
\usepackage{amssymb}
\usepackage[amssymb]{SIunits}
\usepackage{multirow}
\usepackage{bbold}
\usepackage{verbatim}
\usepackage{comment}
\usepackage{caption}
\usepackage{physics}
\usepackage{subcaption}

\definecolor{darkgreen}{HTML}{339933}

\usepackage{scalerel,tikz}
\usetikzlibrary{svg.path}
\definecolor{orcidlogocol}{HTML}{A6CE39}
\tikzset{orcidlogo/.pic={
 \fill[orcidlogocol] svg{M256,128c0,70.7-57.3,128-128,128C57.3,256,0,198.7,0,128C0,57.3,57.3,0,128,0C198.7,0,256,57.3,256,128z};
 \fill[white] svg{M86.3,186.2H70.9V79.1h15.4v48.4V186.2z}
 svg{M108.9,79.1h41.6c39.6,0,57,28.3,57,53.6c0,27.5-21.5,53.6-56.8,53.6h-41.8V79.1z M124.3,172.4h24.5c34.9,0,42.9-26.5,42.9-39.7c0-21.5-13.7-39.7-43.7-39.7h-23.7V172.4z}
 svg{M88.7,56.8c0,5.5-4.5,10.1-10.1,10.1c-5.6,0-10.1-4.6-10.1-10.1c0-5.6,4.5-10.1,10.1-10.1C84.2,46.7,88.7,51.3,88.7,56.8z};
}}
\newcommand\orcidicon[1]{\href{https://orcid.org/#1}{\mbox{\scalerel*{
\begin{tikzpicture}[yscale=-1,transform shape]
\pic{orcidlogo};
\end{tikzpicture}
}{|}}}}

\title{Cosmology before noon with multiple galaxy populations}

\author[a,b,c]{Haruki Ebina~\orcidicon{0000-0002-1080-0955},}
\author[a,b,c]{and Martin White~\orcidicon{0000-0001-9912-5070}}
\affiliation[a]{Department of Physics, University of California, Berkeley, CA 94720, USA}
\affiliation[b]{Berkeley Center for Cosmological Physics, UC Berkeley, CA 94720, USA}
\affiliation[c]{Lawrence Berkeley National Laboratory, One Cyclotron Road, Berkeley, CA 94720, USA}

\emailAdd{ebina@berkeley.edu}
\emailAdd{mwhite@berkeley.edu}

\abstract{
Near-future facilities observing the high-redshift universe ($2<z<5$) will have an opportunity to take advantage of ``multi-tracer'' cosmology by observing multiple tracers of the matter density field: Lyman alpha emitters (LAE), Lyman break galaxies (LBG), and CMB lensing $\kappa$. In this work we use Fisher forecasts to investigate the effect of multi-tracers on next-generation facilities. 
In agreement with previous work, we show that multiple tracers improve constraints primarily from degeneracy breaking, instead of the traditional intuition of sample variance cancellation. Then, we forecast that for both BBN and CMB primary priors, the addition of lensing and LAEs onto a LBG-only sample will gain 25\% or more in many parameters, with the largest gains being factor of $\sim10$ improvement for $f_{\rm EDE}$. 
We include a preliminary approach towards modelling the impact of radiative transfer (RT) on forecasts involving LAEs by introducing a simplified model at linear theory level. Our results, albeit preliminary, show that while RT influences LAE-only forecasts strongly, its effect on composite multi-tracer forecasts is limited. 
}

\begin{document}
\maketitle
\flushbottom

\section{Introduction}
\label{sec:introduction}

The large-scale structure of the Universe arises from primordial fluctuations in space-time in the very early Universe, acted upon by gravitational instability in a dark-matter-dominated Universe over many Gyr \cite{Peacock99,Dodelson03,Dodelson20,Baumann22}.  As such it provides us with a window into fundamental physics and all of the constituents, processes and parameters that influence the evolution of large-scale structure even in subtle ways.  In turn this has led to exciting proposals for next-generation facilities that can make high precision measurements of the quasi-linear modes that contain such a wealth of information \cite{Sailer21,SpecTel,dawson2018cosmic,Blanchard:2021ffq,Schlegel22}.

Much of the focus of these next generation facilities will be measurements of large-scale structure at redshifts above $z\sim 1-2$ in order to take advantage of several simultaneous, advantageous trends \cite{Ferraro22}.  First, probing a wide lever arm in time (redshift) is critical to disentangle evolutionary influences from primordial ones and leads to rotated degeneracy directions that tighten joint constraints.  Second, the larger volume and smaller scale of non-linearity both allow a larger lever arm in scale to break degeneracies between cosmological and nuisance parameters.  Third, the modes are in the linear or quasi-linear regime and carry information from the early Universe before it has been lost to non-linear evolution.

A fundamental limitation to how well we can probe fundamental physics and cosmology comes from the influence of complex astrophysical processes that affect the clustering of objects on the length scales that we observe.  This influence is usually described by a set of bias parameters \cite{Desjacques18} whose form is dictated by the symmetries of our problem (e.g.\ rotational invariance, the equivalence principle, ...).  The ability to break degeneracies between these bias parameters and the fundamental parameters of interest is key to the success of the program.  Such degeneracy breaking is dramatically improved if we have multiple tracers of the same underlying density fluctuations with different (or no) biases, such as multiple galaxy populations (known as ``multi-tracer''; \cite{McDonald09}) or gravitational lensing (of the CMB; \cite{Lewis06,Hanson10}).
Traditionally, multitracer has been considered in the context of `sample variance cancellation', which avoids the payment of sample variance by measuring quantities, such as ratios of biases, at field level. However, for our forecast we will not have high enough number densities for this to be interesting, nor will we present forecasts for parameters that can be measured at field level.

At $z>2$ there are two sets of galaxies that make natural targets for spectroscopic surveys: Lyman break galaxies (LBGs) and Lyman alpha emitters (LAEs).  LBGs are massive, actively star-forming galaxies with a luminosity approximately proportional to their stellar mass.  They are abundant, well studied and well understood \cite{Giavalisco02,Shapley11,Wilson19}.  By contrast LAEs are compact, metal-poor, star-forming galaxies with SFR$\sim 1-10\ M_\odot\mathrm{yr}^{-1}$ that live in much lower mass halos and consequently are expected to have significantly lower bias \cite{Ouchi20}. 
These two classes of targets naturally possess different biases, making them ideal laboratories for the multi-tracer approach.  In addition we have lensing of the CMB, which provides a (projected) tracer of the matter field itself (i.e.\ with no biases) which further serves to break degeneracies.

In this paper we extend the Fisher formalism developed in ref.~\cite{Sailer21} to the case of multiple galaxy populations (plus the CMB lensing already included in \cite{Sailer21}) and forecast the improvement in constraining power that this provides.  The outline of the paper is as follows.
In \S\ref{tracers}, we introduce and characterize the three matter density tracers. In \S\ref{sec:theory} we describe the theoretical framework used for forecasting, including the modelling of biased tracers and the Fisher information formalism. In \S\ref{multitracer} we then investigate the mechanism of multi-tracer cosmology through forecasting of simplified galaxy tracers with varying characteristics. \S\ref{sec:RT} explains our approach to radiative transfer (RT), which potentially poses concerns toward the use of LAEs as tracers. Finally in \S\ref{sec:results} we present our forecasts for two experiment designs, Stage 4.5, motivated by DESI II, and Stage V, and conclude in \S\ref{sec:conclusion} with discussion. 

\section{Tracers of large-scale structure at high redshift} \label{tracers}

In this section we describe the relevant properties of the three tracers used for this study; Lyman alpha emitters (LAEs), Lyman break galaxies (LBGs), and gravitational lensing. For the galaxy tracers we determine the large-scale bias, $b$, and comoving number density, $\bar{n}$, of the objects based on previous astronomical surveys that have probed these tracers at the redshifts of relevance. The distinction between these tracers is made primarily by observational technique than physical properties, causing overlap of these populations in many cases. 

\subsection{Lyman Alpha Emitters} \label{LAEs}

Lyman alpha emitters (LAEs) are young, low-mass, actively star-forming galaxies at high redshift \cite{Dijkstra14, Ouchi20}.  In the astronomical literature the designation is primarily determined by the method of sample selection, and LAEs are normally selected from narrow or intermediate band imaging surveys as galaxies having large rest-frame equivalent width Ly$\alpha$ emission \cite{Garel15, Ouchi20}. 
While practical, this designation complicates the physical interpretation of the population. Although the dominant physical origins of these emissions (e.g.~young massive stars, AGNs, and shock heating with minimal absorption by dust) have been identified \cite{Ouchi20}, a complete understanding of what determines the high Ly$\alpha$ emissivity of LAEs has not yet been obtained. Although we will largely follow the observational designation, from the perspective of cosmology the primary interest will be in the ability to obtain a redshift for the galaxy, which is more closely associated with line flux. There may also be arguments for having detectable continuum emission in order to minimize radiative transfer effects on the galaxy selection.

Due to the method of selection, the characterization of the LAE population for the purpose of cosmological forecasts contains some subtleties and difficulties.  The majority of LAE surveys probe faint populations over relatively small areas of sky, whereas the ideal objects for wide-field spectroscopy tend to be brighter and not well-represented in such surveys. 
Pure narrow-band selection is also not likely to be an efficient means of finding galaxies over large cosmological volumes due to the limited comoving depth each filter provides. For this reason future cosmological surveys will likely rely on targets selected, at least in part, from medium-band imaging.  The extent to which the clustering properties of medium band selected objects differ from those selected in narrow bands remains to be quantified. This, along with varying observational definitions of LAE across literature (e.g.\ whether to include an equivalent width cut, and if so of what value, or whether to include objects with no continuum emission), means our estimates have an element of uncertainty. It is to be expected that differences in sky area coverage, redshift range, and instrumentation between existing surveys and next-generation cosmological facilities indicate that galaxy populations will likely not be identical between them.

The currently ongoing ODIN (One-hundred-$\rm deg^2$ DECam Imaging in Narrowbands) survey \cite{Lee24} is amongst the widest of existing LAE surveys (100$\rm deg^2$), though even this is at least a factor of 50 less volume than the samples considered in this work.  Unfortunately, the clustering properties of the ODIN sample have not been published, so we primarily rely on previous surveys of even smaller areas, such as the Slicing COSMOS 4K (SC4K) survey \cite{Sobral18,Khostovan19} at $\sim 2{\rm deg}^2$. While the SC4K covers significantly less area than ODIN, it provides a wider range of redshifts through the use of multiple narrow- and medium-band filters.  Though the samples are still not an ideal match for the cosmological samples we envision, they represent our current best observational constraints on such populations.

\begin{figure}
    \centering
    \includegraphics[width=\textwidth]{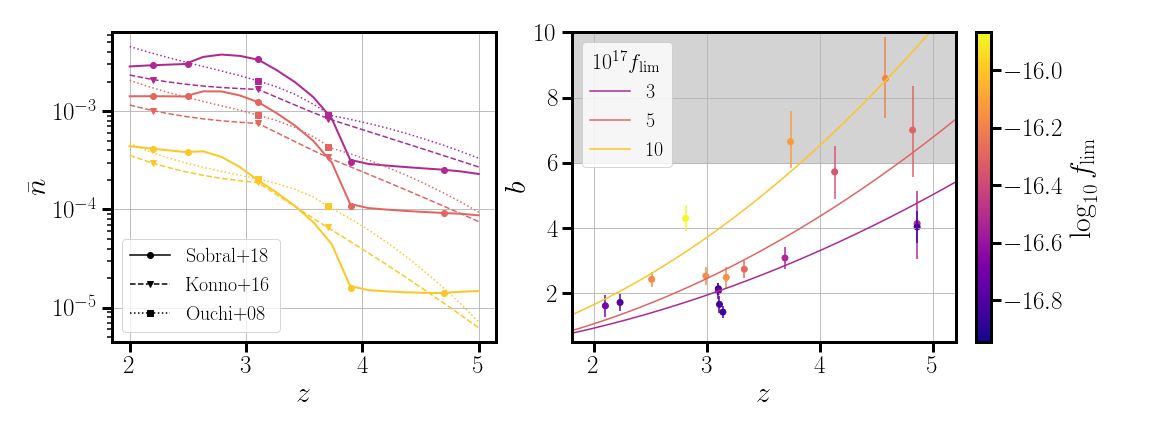}
    \caption{Left: 3D number density ($\bar{n}$) of LAEs found using Schechter function fits to LF from previous studies \cite{Sobral18,Konno16,Ouchi08}. The point markers refer to the values reported in the original studies, while the lines use interpolations of Schechter LF parameters. We see that although there is rough consistency across the literature, a `factor of two' difference is not uncommon. We take the LF parameters from SC4K \cite{Sobral18} as our fiducial model in what follows.
    Right: Large scale bias $b$ of LAEs reported across the literature with various redshift and line flux limits \cite{Khostovan19,Bielby16,Ouchi10,Shioya09,Hao18,Gawiser07,Guaita10}. The solid lines show the fit of Eq.\ (\ref{bias_fit}) for $f_{\rm lim}=3$, 5 and $10\times 10^{-17}\,\text{erg}\,\text{cm}^{-2}\,\text{s}^{-1}$. The region $b\ge 6$ is shaded, as populations with such high bias are not traditionally used for cosmology.}
\label{fig:LAEnb}
\end{figure}

The two key parameters for our forecasts will be the abundance and large-scale clustering of LAEs, as determined by the 3D number density ($\bar{n})$ and large-scale bias ($b$).  In order to determine these we make use of published studies \cite{Khostovan19, Sobral18, Konno16, Ouchi08, Guaita10, Bielby16, Shioya09, Hao18, Gawiser07}, which we discuss further below. The final inputs are shown in Figure \ref{fig:LAEnb}.  

LAE luminosity functions (LF) are frequently quoted in terms of the Ly$\alpha$ luminosity and fit with the Schechter functional form \cite{Schechter76}.  While the uncertainties in the three parameters of the Schechter function are often covariant, this information is not usually provided which makes assessing uncertainties in $\bar{n}$ more difficult. Thus, we choose to compare different surveys as an estimate of uncertainty.  This has the advantage of also including sample variance due to survey volume, though it unfortunately mixes differences in observational selection into the uncertainty.

Assuming a Schechter LF the number density, $\bar{n}$, of objects brighter than some minimum luminosity, $L_{\rm lim}$, can be written in terms of three parameters ($\phi^*$, $L^*$ and $\alpha$) as
\begin{equation}
    \bar{n}(z) = \ln{10} \int^\infty_{\log_{10}L_{\rm lim}} \ \phi^\star \left(\frac{L}{L^*}\right)^{1+\alpha} e^{-L/L^*} d\log_{10} L
\end{equation}
where $L$ is the Ly$\alpha$ line luminosity. As it ties more directly to observational `cost', we will typically use the line flux ($f$) instead of luminosity, the two are related by $L = 4\pi f\, D_L^2(z)$, with $D_L$ the luminosity distance.

In Figure \ref{fig:LAEnb}, we compare number densities using various LF parameters from different surveys \cite{Sobral18,Ouchi08,Konno16}. While there is rough consistency, there are frequently factors of $\sim 2$ between curves, suggesting our estimates of $\bar{n}$ are likely uncertain `at the factor of two level'.  We will investigate the impact of this later in \S\ref{sec:results}. We adopt the LF from the SC4K survey \cite{Sobral18} as our fiducial model. Since the LF-derived number densities reflect the total number of galaxies at a given line flux limit after correcting for incompleteness, we further include an estimated 50\% selection and redshift efficiency. 

For our forecasts we will also need to understand the large-scale bias of these populations, which is more complex than the number densities.  Existing surveys quote a range of different measures (see e.g.\ Fig.~3 of ref.~\cite{Khostovan19} for a recent compilation).  
The clustering strength in LAE surveys is often quoted in terms of a correlation length, $r_0$, assuming a power-law (real-space) correlation function:
\begin{equation}
    \xi_{gg}(r) = \left(\frac{r}{r_0}\right)^{\gamma}
\end{equation}
and the power-law index $\gamma$ is often held fixed (typically to $-1.8$) in the fit.  Given that astronomical LAE surveys typically have discrete narrow- or medium-band selection over a small field, in practice $r_0$ is obtained by fitting the angular correlation function, $w(\theta)$, to a power-law.  Ignoring redshift-space distortions (RSD) the projection of an isotropic 3D correlation, $\xi(r)$, to 2D would give $w(\theta)\propto \theta^{\beta}$ with $\beta=\gamma+1$.  The presence of redshift-space distortions leads to appreciable changes in this relation on scales comparable to the depth of the shell in comoving distance, an effect that is usually neglected in the literature.

Given the limited information available, we will convert the reported $r_0$ values to large-scale biases by working to linear order in bias, in which case
\begin{equation}
    b^2 = \frac{\xi_{gg}}{\xi_{mm}}
\label{eqn:biasdef}
\end{equation}
Again, any RSD contribution to the linear power-spectrum is ignored assuming that the constraints are primarily from scales much smaller than the depth of the survey. A subtlety involved here is that while we are using the linear approximation for the bias, the power-law approximation for $\xi_{gg}$ assumes nonlinear growth which is normally associated with non-linear, scale-dependent bias. Although this poses an inconsistency in our approach, considering higher order biases or re-investigating raw data to find the exact form of $\xi_{gg}$ goes well beyond the scope of this work and likely will not alter the conclusions of this study significantly.  It would be worthwhile to refine these estimates when a new analysis of the clustering of LAEs over wide areas becomes available.

This method for estimating the bias requires us to make a number of choices.  We evaluate $b$ by computing the ratio in Eq.~\ref{eqn:biasdef} at a fixed scale, $r$.  Since the above estimation assumes that the power-law approximation of $\xi_{gg}$ is valid we choose an angular scale of $\theta\sim 235^{\prime\prime}$. The validity of the power-law approximation at this scale across surveys and redshifts can be confirmed through inspection of the power-law fit to the angular correlation in the literature \cite{Guaita10,Khostovan19}.  This scale corresponds to $r\simeq 5\,h^{-1}\,\text{Mpc}$ at $z=3$, and $5\,h^{-1}$Mpc is large enough to be in the quasi-linear regime for $z>2$ while being smaller than the typical survey depth to reduce the impact of RSD.  The numerator is estimated from the power-law fit, while for the denominator we use the Hankel transform of the HaloFit \cite{Takahashi12} power spectrum as computed by CLASS \cite{CLASS} which is approximately a power-law on these scales.

We utilize clustering measurements from various surveys in the literature \cite{Khostovan19,Bielby16,Ouchi10,Shioya09, Hao18,Gawiser07,Guaita10}. Note that we adopt $r_0$ values from \cite{Khostovan19} for all surveys, as they have reported the values from the literature after normalizing for some methodology differences.
We estimate $b$ as described above for a range of redshifts ($z$) and flux limits ($f_{\rm lim})$ and the results are shown in Fig.~\ref{fig:LAEnb}. For convenience we fit these results with
\begin{equation}
    b(f_{\rm lim},z) = A(f_{\rm lim}) (1+z) + B(f_{\rm lim}) (1+z)^2 
    \label{bias_fit}
\end{equation}
where
\begin{align}
    A(f_{\rm lim}) &= 0.457-1.755(\log_{10}f_{\rm lim}+17)+0.720(\log_{10}f_{\rm lim}+17)^2 \\
    B(f_{\rm lim}) &= 0.012+0.318(\log_{10}f_{\rm lim}+17)+0.043(\log_{10}f_{\rm lim}+17)^2
\end{align}
are both quadratic functions of $\log_{10}{f_{\rm lim}}$, with the flux measured in $\text{erg}\,\text{cm}^{-2}\,\text{s}^{-1}$ by standard convention. The shape of this fit is motivated by a similar fit to the LBG bias in ref.~\cite{Wilson19}.  Since at these high redshifts the linear growth rate $D(z)\propto (1+z)$ the $(1+z)$ term corresponds to co-evolution \cite{Tegmark98} while the $(1+z)^2$ term is its first order correction. Ultimately, however, this is simply a convenient fit to the observed results.  The solid lines in Fig.~\ref{fig:LAEnb} show the fit for $f_{\rm lim}=3$, 5 and $10\times 10^{-17}\,\text{erg}\,\text{cm}^{-2}\text{s}^{-1}$. Note that while we choose to model $b$ as a function of $f_{\rm lim}$, which can be derived from $L_{\rm lim}$, it is common for surveys to report the limiting luminosity in terms of the median luminosity $L_{\rm med}$. The conversion from $L_{\rm med}$ to $L_{\rm lim}$, assuming a Schechter luminosity function, is described in Appendix \ref{luminosity}.

\subsection{Lyman Break Galaxies}

\begin{figure}
    \centering
    \includegraphics[width=\textwidth]{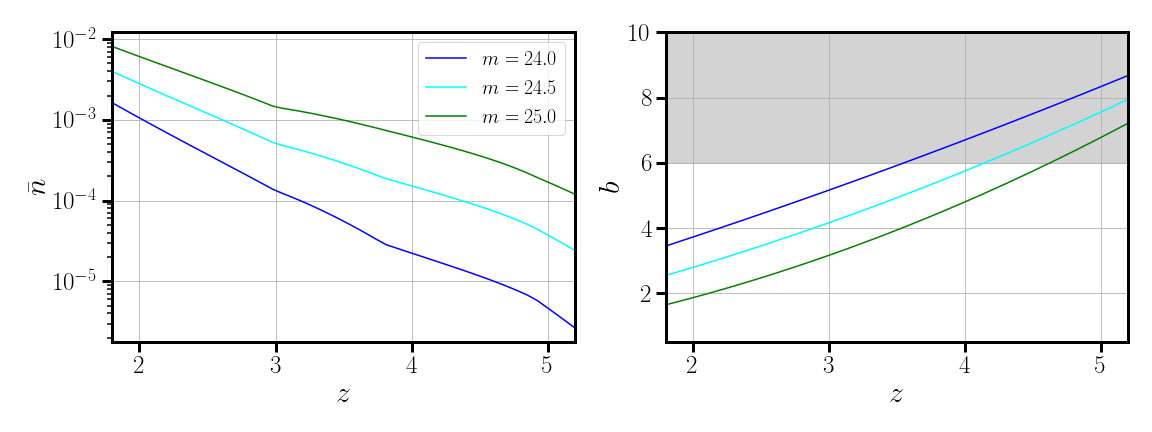}
    \caption{Left: Number density $\bar{n}$ of LBGs using Schechter function fits to the UV LF \cite{Wilson19}, using fit parameters reported from preceding studies \cite{Reddy08,Malkan17,Ono17}. Right: Large-scale bias $b$ of LBGs using a fit function $b(m,z)$ in Eq.~\ref{b(m,z)} \cite{Wilson19}. The region $b\ge6$ is shaded to indicate that these high bias tracers are traditionally not used for cosmology.}
    \label{fig:LBGnb}
\end{figure}

Lyman break galaxies (LBGs) \cite{Wilson19,Giavalisco02,Shapley11} are a set of massive, actively star-forming galaxies at high redshift. The population is typically selected through a `dropout' technique in which the selection targets the distinct drop in flux density $F_\nu$ that occurs at 912\AA\ (the Lyman break) due to absorption by stellar atmospheres and interstellar medium on top of an otherwise flat spectrum. For the redshifts of our interest, the spectra bluewards of 1216\AA\, is further suppressed due to Lyman-series blanketing \cite{Madau95}. Importantly, selection for this drop yields physically similar galaxies across redshift. 
Each selection sample is typically named after the filter that is used to detect this drop. For instance, a $u$-dropout sample will have a lack of flux in the $u$ band compared to significant flux in redder bands, corresponding to galaxies at $z\approx 3$. The typical width for each sample is $\Delta z\approx0.7$.
Similar to LAEs, the number density $\bar{n}$ and large scale bias $b$ are our key parameters, allowing us to use a nearly identical approach for parameter estimation. The results are summarized in Figure \ref{fig:LBGnb}. 

The Schechter formalism is once again useful to predict the number density. Here, we will be using the UV ($\approx 1500$\AA\, rest-frame) LF with absolute magnitude $M_{\rm UV}$
\begin{equation}
    dn(M_{\rm UV}) = \left(\frac{\ln{10}}{2.5}\right) \phi^* 10^{-0.4(1+\alpha)(M_{\rm UV}-M_{\rm UV}^*)} \exp{-10^{-0.4(M_{\rm UV}-M_{\rm UV}^*)}} dM_{\rm UV}
\end{equation}
with the LF parameters from Table 3 of ref.~\cite{Wilson19} (in turn based on refs.\ \cite{Reddy08,Malkan17,Ono17}). The absolute magnitude can be converted to apparent magnitude, $m$, using the appropriate (near-zero) $k$-correction \cite{Wilson19}
\begin{equation}
    M_{\rm UV} = m - 5\log_{10}\left(\frac{D_L(z)}{10\text{pc}}\right)+2.5\log_{10}(1+z)+\underbrace{m_{\rm UV}-m}_{\approx0}
\end{equation}
We will use this to model (apparent) magnitude limited surveys.

The LBG biases are calculated using the same power-law approximation of correlation functions as in LAEs. Ref.\ \cite{Wilson19} fit a function $b(m,z)$ to results from the literature
\begin{equation}
    b(m,z) = A(m)(1+z) + B(m)(1+z)^2 
    \label{b(m,z)}
\end{equation}
where $A(m)=-0.98(m-25)+0.11$ and $B(m)=0.12(m-25)+0.17$. Again, while this fitting model has some physical basis on the co-evolution approximation of galaxies, it is ultimately a useful fit to the data. 

\begin{table}[]
    \centering
    \begin{tabular}{c||c|c|c|c|c|c}
        $z$ & 2.2 & 2.5 & 3.1 & 3.9 & 4.7 & 5.4 \\
        \hline
        LAE $f_{\rm lim}^*$ & 13 & 11 & 5.6 & 5.1 & 5.4 & 6.9 \\
        \hline \hline
        $z$ & 2 & -- & 3 & 3.8 & 4.9 & 5.9 \\
        \hline
        LBG $m^*$ & 24.2 & -- & 24.7 & 25.4 & 25.5 & 25.8 \\
    \end{tabular}
    \caption{Characteristic luminosity of Schechter function fits to the LAE LF \cite{Sobral18} and LBG LFs \cite{Wilson19} (in turn based on \cite{Reddy08,Malkan17, Ono17}), converted to line flux (in units of $10^{-17}\,\text{erg}\,\text{cm}^{-2}\,\text{s}^{-1}$) and apparent magnitude, respectively. This can be used to gauge what the optimal limiting luminosity is for experiments, as the LF will transition from an exponential to a power-law fainter than these characteristic luminosities and provide diminishing returns in galaxy counts. 
    }
    \label{tab:L*}
\end{table}

The summary of the characteristic luminosities for LAEs and LBGs, shown in Table \ref{tab:L*} provide us with an intuitive limit on where our gains are maximized when pushing to fainter luminosities.  Starting at high luminosities, the number densities increase exponentially as we push fainter until we reach the ``knee'' at the characteristic luminosity ($L^\star$), where the increase turns to a power-law. Near-term surveys will be able to reach such regimes. 

\begin{table}[]
    \centering
    \begin{tabular}{c|c|c|c|c|c|c|c}
        \multirow{2}{6em}{Experiment}& \multicolumn{2}{c|}{$2.4<z<3.2$} & \multicolumn{2}{c|}{$3.2<z<4.4$} & \multicolumn{2}{c|}{$N\, [1/\text{deg}^2]$} & \multirow{2}{3em}{$f_{\rm sky}$} \\
        \cline{2-7}
        & $f_{\rm lim}$& $m$ & $f_{\rm lim}$& $m$ & LAE & LBG & \\
        \hline
        Stage 4.5 & 10 & 24.5 & -- & -- & 489 & 1237 & 0.1212 \\
        Stage V Plan A & 10 & 24.5 & 10 & 24.5 & 611 & 1716 & 0.4\\
        Stage V Plan B & 10 & 24.5 & 5 & 24.5 & 1294 & 1716 & 0.3092 \\
        Stage V Plan C & 10 & 24.5 & 10 & 25.0 & 611 & 2959 & 0.2607
    \end{tabular}
    \caption{Experiment designs considered for this work. $f_{\rm lim}$ and $m$ refer to the limiting flux of LAE samples and limiting apparent magnitude of LBG samples, respectively. The units of $f_{\rm lim}$ are $10^{-17}\text{erg}/\text{cm}^2/\text{s}$. The Stage 4.5 design is strongly motivated by the proposed DESI II design.}
    \label{tab:experiments}
\end{table}

For this work, we will consider a `Stage 4.5' design motivated by the DESI II design with $2.4<z<3.2$, and a `Stage V' design with $2.4<z<4.4$. The redshift ranges are inspired by existing survey designs \cite{Schlegel22} and ongoing work by DECam probing LAEs up to $z\sim4.5$ \cite{Lee24}. Both designs are constrained by the limiting luminosities for LAEs and LBGs. While Stage 4.5, with a closer timeline, is specified in luminosity limits, we will consider multiple scenarios of Stage V luminosity limits to perform a crude optimization. The optimization will be done by assuming both a fixed total integration time for the experiment and a fixed individual integration time per galaxy, regardless of flux, magnitude, or galaxy type. This is equivalent to fixing the total number of galaxies observed and we achieve this by tuning $f_{\rm sky}$ for each Stage V design to counteract change in number density by the choice of luminosity limits. 
The experimental designs are shown in Table \ref{tab:experiments}. Note that these designs are exploratory by nature, as neither the capabilities of next-generation facilities nor tracer characteristics are determined and will benefit from a forecast with broad considerations rather than precise optimization.
A comparison against Table \ref{tab:L*} shows that for a Stage 4.5 survey with $f_{\rm lim}=10\times 10^{-17}$, $m=24.5$, and $z\lesssim 3$, the luminosity limits are near the characteristic luminosities with the exception of LAEs at $z=3.1$. 

\begin{figure}
    \centering
    \includegraphics[width=\textwidth]{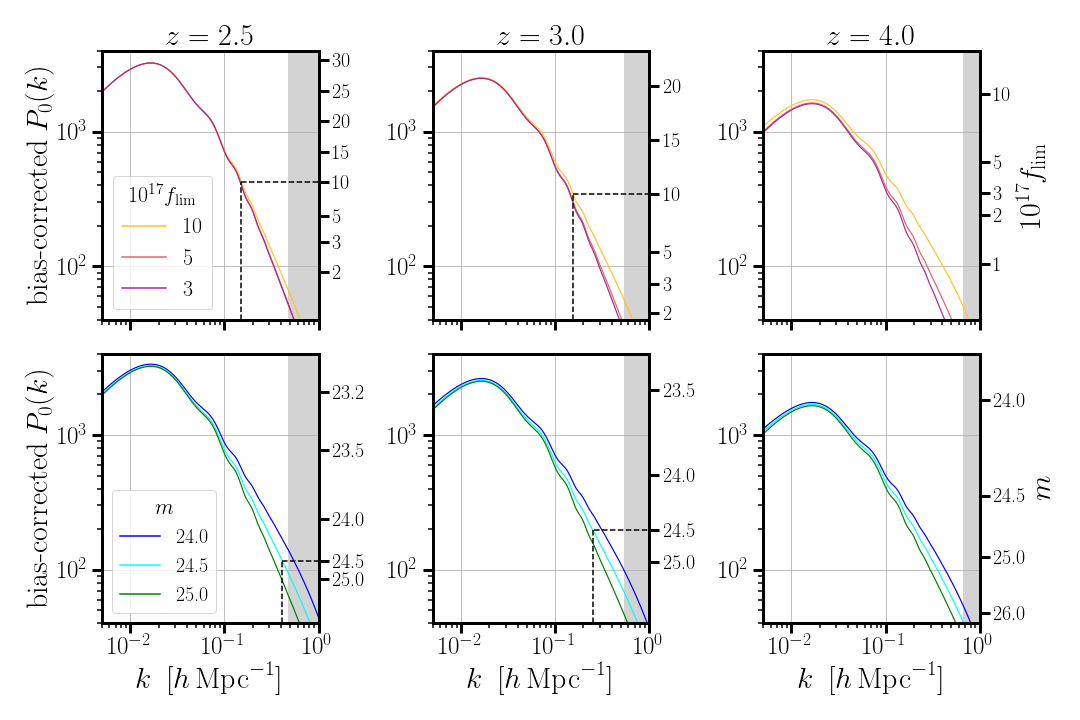}
    \caption{Monopole of the signal ($P-P_{\rm stoch}$) against the shot noise at various luminosity limits (right axis) for LAEs (top) and LBGs (bottom). The $k$ where signal equals shot noise reflects where we enter a shot noise dominated regime ($\Bar{n} P$=1). 
    The bias-correction simply refers to dividing out the monopole Kaiser factor, $b^2+2bf/3+f^2/5$, so that the large-scale clustering is independent of magnitude.  The spread in the curves at high-$k$ is due to scale-dependent bias, which we model as described in \S\ref{sec:theory}. The dotted lines illustrate where the fiducial Stage 4.5 survey become shot noise dominated.
    }
    \label{fig:P vs SN}
\end{figure}

A more quantitative measure of the strength of our signal is where in $k$-space we become shot noise limited in the monopole, as shown in Figure \ref{fig:P vs SN}. For the Stage 4.5 survey, the LAE signals become shot noise limited at $k\simeq 0.15\,h\,\mathrm{Mpc}^{-1}$ and LBGs at $k\simeq 0.3-0.4\,h\,\mathrm{Mpc}^{-1}$. Furthermore, if we extend these luminosity limits to $z=4$, we are totally shot noise dominated for LAEs. 
Notice that the magnitude limits on the right axis do not go to 26 for $z=2.5$ and 3.0. This is because the bias-corrected shot noise $[\bar{n}(b^2+2bf/3+f^2/5)]^{-1}$ increases at fainter magnitudes due to the decrease in bias over-powering the increase in number density as we push fainter than the corresponding characteristic magnitudes of the LBG Schechter function fit.

\subsection{CMB Lensing}

\begin{figure}
    \centering
    \includegraphics[width=.8\textwidth]{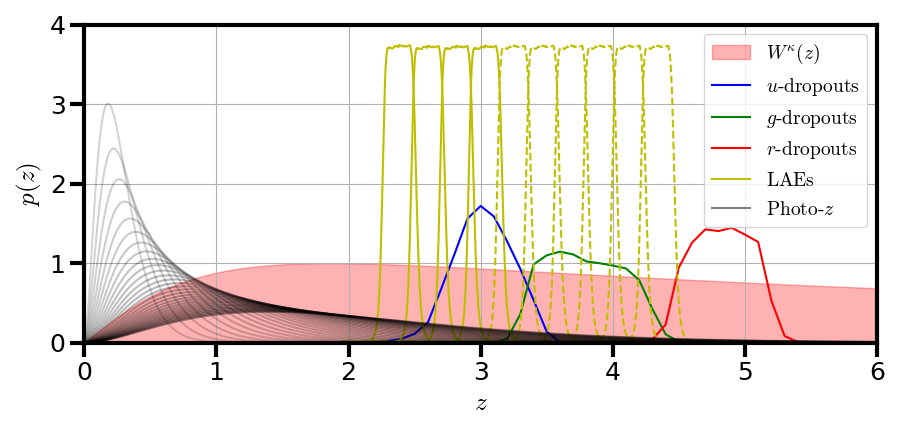}
    \caption{The normalized redshift distribution of the matter tracers considered for this work. The colored curves indicate the normalized redshift distribution of galaxy samples \cite{Wilson19,Ono17,Arjun} observed by spectroscopic experiments in Table \ref{tab:experiments}. The LAE filter limits for Stage 4.5 redshifts are inspired by an ongoing DECam medium-band survey \cite{Arjun} and extrapolated for higher redshifts. Note that the filters of LAE samples are not identical the redshift bins of the forecast. The light red region indicates the CMB lensing kernel $W^\kappa(z)\propto W^\kappa(\chi)/H(z)$ normalized to 1 at maximum. Also included is a common approximate distribution of magnitude limited samples in photometric surveys \cite{Wilson19,LSST,LSST09} that we do not forecast for in this work. The magnitude limits are $i_{\rm lim}={20,20.5,\dots 35}$, with the higher magnitude limit samples extending to higher redshift. }
    \label{fig:dNdz}
\end{figure}

We will also include CMB lensing $\kappa$ as a third tracer of the matter density. The aspect of lensing that makes it a particularly interesting addition to the set of tracers is that it introduces fewer nuisance parameters than a galaxy tracer; it is an unbiased tracer of matter, with no shot noise. The high redshifts of near-term experiments further motivate the use of lensing, as the CMB lensing kernel peaks near these redshifts. On the other hand, we will not be including cosmic shear as a tracer, as this will require galaxy sources at extremely high redshift. Refer to Figure \ref{fig:dNdz} for the full redshift overlap between density tracers considered for this work. 

We consider Simons Observatory (SO) \cite{SimonsObs} with $f_{\rm sky}=0.4$ as the CMB experiment for our forecasts and use the noise $N^\kappa_\ell$ obtained from the \href{https://github.com/simonsobs/so_noise_models/blob/master/LAT_lensing_noise/lensing_v3_1_0/nlkk_v3_1_0deproj0_SENS2_fsky0p4_it_lT30-3000_lP30-5000.dat}{SO noise calculator} as in ref.~\cite{Sailer21}. The calculator uses an iterated minimum variance (MV) quadratic estimator and MV internal linear combination of both CMB temperature and polarization data for calculation, and includes atmospheric effects \cite{Sailer21}.

\section{Theory and formalism}
\label{sec:theory}

In this section we outline the theoretical models we employ and the forecasting formalism we adopt, which is based on the Fisher matrix and is substantially similar to that used\footnote{\hyperlink{https://github.com/NoahSailer/FishLSS}{https://github.com/NoahSailer/FishLSS}} in ref.~\cite{Sailer21}.  We take $\Lambda$CDM as our fiducial cosmology with parameters given in Table \ref{tab:fiducial}.

\subsection{Model}

As in ref.~\cite{Sailer21}, our theoretical model is based upon 1-loop perturbation theory with a general, quadratic bias model.  This model has been shown to be accurate (at the relevant scales) well beyond the demands of current and next-generation surveys \cite{Chen20c} while including scale-dependent bias and mode-coupling due to quasi-linear evolution that improve the reliability of the forecasts.  It has been applied to earlier surveys in ref.~\cite{White22,Chen22a,Chen22b} and is currently being evaluated for use in the DESI and Euclid surveys \cite{MARK}.  For further details of the model see \cite{Chen20a,Chen20b,Chen20c}.

\begin{figure}
    \centering
    \includegraphics[width=\textwidth]{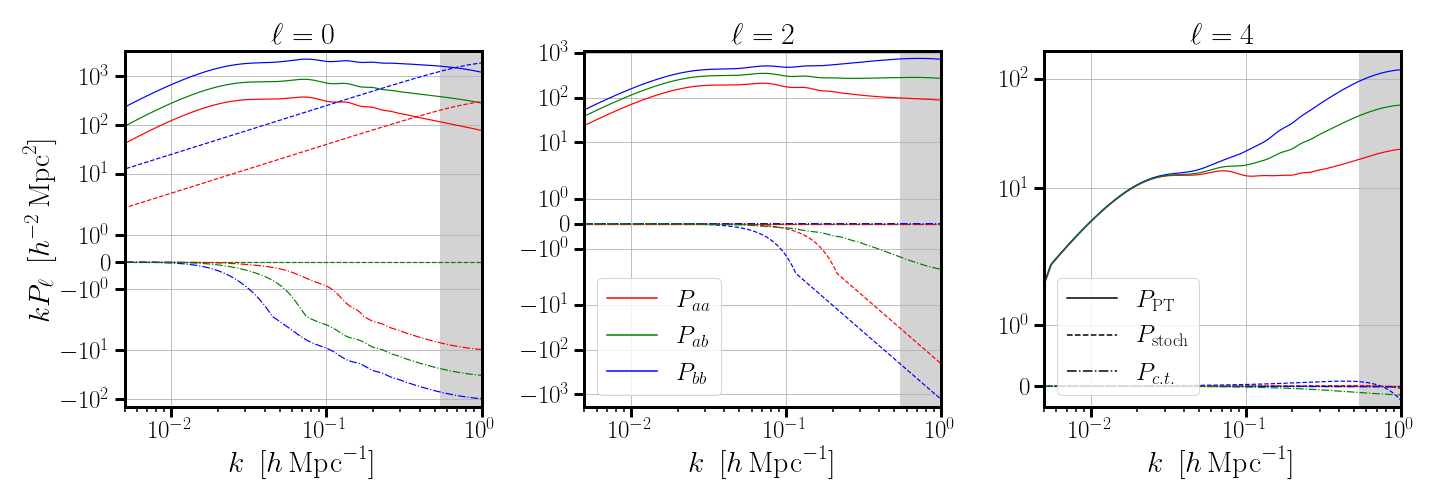}
    \caption{The contributions to the power spectra from each component in Eq. (\ref{general_spectra}), where $b_a=2$, $b_b=5$, $z=3$, and $P_{\rm PT}$ refers to the sum of quadratic bias component spectra $\sum_{X,Y} b_X b_Y P_{XY}(k,\mu)$. The columns refer to the monopole, quadrupole, and hexadecapole of each component. Red lines show the $aa$ auto-spectra, green lines the $ab$ cross-spectra, and blue lines the $bb$ auto-spectra. The shaded region reflects the $k_{\rm nl}$ limit at $z=3$.
    }
    \label{fig:show_spectra}
\end{figure}

Within this formalism, the auto- or cross-spectra for our tracers can be written as a linear combination of different `basis' spectra multiplied by bias coefficients, plus contributions from counterterms and stochastic terms:
\begin{equation}
    P(k,\mu) = \sum_{X,Y} b_X b_Y P_{XY}(k,\mu) + P_{c.t.}(k,\mu) + P_{\rm stoch}(k,\mu)
    \label{general_spectra}
\end{equation}
where $X, Y\in \{m,\delta_L,\delta_L^2,s_L^2,\nabla^2\delta_L,\cdots\}$ label the component spectra, $P_{c.t.}$ are counterterms arising from small-scale physics not directly included in the model, and $P_{\rm stoch}$ are stochastic terms such as shot noise and fingers of god. 
Figure \ref{fig:show_spectra} shows the contribution of each component in a sample case of constant biases, $b_a=2$ and $b_b=5$, constant number density, $\Bar{n}=(100 b^2)^{-1}$, and no correlation between the shot noise in the two samples (these assumptions correspond to $\Tilde{N}=100$ and $f_{\rm over}=0$, as introduced later in this section). 
Note how the scale and angle dependence of each component spectra differ, allowing us to potentially break degeneracies between parameters. 

\subsection{Parameters}

The parameters that we use can be categorized into `global' and `local' parameters. The global parameters are those that are common to all redshifts, such as cosmological parameters. The local parameters are those that are assigned for each redshift that we consider. The perturbation theory parameters that we describe here ($b_i$, $N_i$, $\alpha_i$) are examples of local parameters. For a complete list of parameters, their categorization (global or local), and their fiducial values, refer to Table \ref{tab:fiducial}.

The bias parameters $b_X$ contribute to the spectra are organized in terms of the dependence of the galaxy overdensity field $\delta_g$ on the matter field $\delta$ and shear $s_{ij}$. Up to one-loop, the bias terms that we include in our model are 
\begin{equation}
    \delta_g = b_1 \delta + \frac{b_2}{2}\delta^2 + b_s s^2
\end{equation}
where $s^2=(s_{ij})^2$ \cite{Chen20a}.  The cubic bias terms are nearly degenerate with our counter- and stochastic terms and so do not need to be included explicitly. This expression corresponds to the parametrization used in Eulerian perturbation thoery (EPT). Each Eulerian bias can be related to the Lagrangian biases $b_X^L$ as \cite{Desjacques18,Chen20a}
\begin{equation}
    b_1 = b^L_1+1, \quad b_2 = b^L_2 + \frac{8}{21}b^L_1, \quad b_s = b^L_s - \frac{2}{7}b^L_1 \quad .
\end{equation}
The fiducial values for each parameter is dependent on that of $b_1$, which we have estimated for both LAEs and LBGs in \S\ref{tracers}. For $b_2$, we use the Sheth-Tormen mass function \cite{Sheth99} and peak-background split model, for which \cite{ColKai89,White14} 
\begin{align}
    b^L_1 &= \frac{1}{\delta_c} \left[ a\nu^2 -1 + \frac{2p}{1+(a\nu^2)^p}\right]\\
    b^L_2 &= \frac{1}{\delta_c^2}\left[ a^2\nu^4 -3a\nu^2 + \frac{2p(2a\nu^2+2p-1)}{1+(a\nu^2)^p}\right]
    \label{sheth-tormen}
\end{align}
with $\delta_c=1.686$, $a=0.707$ and $p=0.3$. This is shown to agree well with N-body simulations for $b_2$ \cite{Abidi18} and thus motivates a reasonable fiducial value for our samples. Furthermore, we have confirmed that our results are qualitatively invariant to our choice of $b_2$. For $b_s$, we use the coevolution approximation, which assumes $b^L_s=0$. While this does not agree as closely with simulations as our choice for $b_2$, simulations suggest that for of our samples we expect $|b^L_s|\lesssim 1$ \cite{Abidi18}, and the differences have little impact on our forecasts. 

Our model for the counterterm consists of parameters $\alpha_{2n,a}$, such that 
\begin{equation}
    P_{a}\supset\alpha_{2n,a}\, \mu^{2n}(k/k_\star)^{2}P_{cb}
\end{equation}
for each auto-spectrum, with $k_\star=1\,h\,\text{Mpc}^{-1}$. We inherit the fiducial values of galaxy $\alpha$'s from ref.~\cite{Sailer21}, where $\alpha_0$ values are determined using fits towards Halofit simulations and higher order parameters are set to 0 for convenience. For $\alpha_{0,\kappa}$, we obtain fiducial values by simply fitting $P_{HF}(z)=P_{\rm lin}(1+\alpha_{0,\kappa}k^2)$. 
Cross-spectra counterterm contributions are fully determined by those of the auto-spectra, thus we do not have degrees of freedom for $\alpha_{ab,a\ne b}$, for example. A detailed discussion of cross-spectrum counterterms is provided in Appendix \ref{sec:counterterms}. 

For the stochastic terms, we use a set of parameters $N_{2n,ab}$, such that
\begin{equation}
    P_{ab}\supset N_{2n,ab}(k\mu)^{2n}
\end{equation}
Our fiducial values of the parameters are survey dependent for $n=0$ or 1, and zero for higher orders. Note that $n=0$ is the ``shot noise term'', and hence our fiducial value is $N_{0,aa}=1/n_a$ for auto-spectra. Extending this model to cross-spectra, we let $N_{0,ab}=f_{\rm over}/\sqrt{n_a n_b}$ fiducially for $a\ne b$, where $f_{\rm over}\in[0,1]$ is a measure of how much correlation we have between the tracers' shot noise (e.g.\ to what degree the populations occupy the same halos). As the fiducial value for $N_2$, we set $N_{2,ab}=-N_{0,ab}\sigma_v^2$, with $\sigma_v=(1+z)(100{\rm km/s})/H(z)$ being the typical velocity dispersion in comoving units \cite{Sailer21}. Notice that the stochastic parameters are the only new degrees of freedom introduced when including the cross-spectra, motivating the use of all available spectra. 

Our $k_{\rm max}$ is set by the non-linear scale. Unless explicitly stated, we will use $k_{\rm max}(z)=k_{\rm nl}(z)=\Sigma^{-1}(z)$ where $\Sigma$ is the rms displacement in the Zel'dovich approximation. We take $k_{\rm min}=0.003\,h\,\text{Mpc}^{-1}$, motivated by the longest mode that fit in our surveys, i.e.\ $k_{\rm min}\sim 2\pi/V^{1/3}$.  Our results are in general not very sensitive to $k_{\rm min}$. 

\subsection{2D power spectrum}

To include CMB lensing in our analysis, we must also have a method to calculate the 2D power-spectra, where we can cross-examine the LSS and CMB data on equal footing. 
The 2D power-spectra can be calculated from the 3D power-spectra using the Limber approximation,
\begin{equation}
    C_\ell^{ab} = \int_0^{\chi_*} d\chi \frac{W^a(\chi)W^b(\chi)}{\chi^2} P_{a b}\left(k_\perp=\frac{\ell+1/2}{\chi},k_\parallel=0\right)
\end{equation}
where the $a$ and $b$ are available tracers, either $\kappa$ or galaxies. $\chi_*$ is the comoving distance to the surface of last scattering and $W$'s are the lensing kernels
\begin{equation}
    W^\kappa(\chi)=\frac{3}{2}\Omega_m H_0^2(1+z)\frac{\chi(\chi_*-\chi)}{\chi_*} \quad\text{and}\quad W^{g_a}(\chi)\propto H(z)\frac{dN_a}{dz}
\end{equation}
where $W^g$ is normalized such that $\int_\chi W^g(\chi)=1$. Note that we take the lowest order correction to the Limber approximation $\ell\rightarrow\ell+1/2$, improving our precision to $\mathcal{O}(\ell^{-2})$ \cite{LovAfs08,Sailer21}. We do not marginalize over higher order stochastic and counter-term parameters for the 2D power-spectra, as the Limber approximation limits us to $\mu=0$. 
For the lensing-lensing auto-spectrum, the $\ell_{\rm max}$ is limited not by $k_{\rm nl}$, but by a stricter constraint from baryonic feedback, which limits us to $\ell<500$. For the rest of the spectra, we translate our $k_{\rm max}$ constraints into $\ell_{\rm max}$. 

\subsection{Fisher formalism}

The Fisher matrix formalism \cite{frieden98} allows us to forecast the parameter constraints on an assumed underlying model, given a experimental configuration.  It is widely used to understand the impact of survey design choices and parameter degeneracies on the inference of cosmological parameters \cite{Tegmark97b,Vogeley96,White09,Chen19a,Sailer21,EuclidFisher}.
The Fisher matrix can be calculated as 
\begin{equation}
    F_{ij}=\sum_{X,Y}\int \frac{k^2\, dk\, d\mu}{2\pi^2}
    \ \frac{\partial P_{X}}{\partial\theta_i} \mathbf{C}^{-1}_{XY} \frac{\partial P_{Y}}{\partial \theta_j}
\end{equation}
where $X,Y$ label available auto- and cross-spectra, $\theta_i$ are parameters, $\mathbf{C}^{-1}_{XY}=(\mathbf{C}^{-1})_{XY}$ and the covariance matrix $\mathbf{C}$ is
\begin{equation}
    \mathbf{C}_{abcd}=P_{ac}P_{bd}+P_{ad}P_{bc}
\end{equation}
where the subscripts represent the samples used to calculate spectra. Note that the power-spectra ($P_X$, $P_{ab}$) above include the counter and stochastic terms. We compute these matrices using the FishLSS code, which was originally developed for single-tracer forecasts \cite{Sailer21} and extended to multitracer forecasts for this work. In the development process we have confirmed that the new version of the code reproduces previous results in single-tracer cases.
Without full consideration of multi-tracer effects, one would calculate the Fisher matrix with $X,\,Y$ summing over only the auto-spectra and thus lose information gained from the cross-spectra. 

For a general galaxy-galaxy-lensing sample, the formalism is modified slightly. While the Fisher matrix earlier still holds, one must compute the matrix using the angular power-spectra $C_\ell^{X}$ as well. One obtains the full set of information by combining the Fisher matrices from the 3D power-spectra and 2D power-spectra. By restricting the full-shape signal to $k_\parallel>10^{-3}$, there will be no covariance between the full-shape and the angular spectra that probe $k_\parallel\sim0$. This will allow for the Fisher matrices to simply add together \cite{Sailer21, frieden98}.

The covariance matrix for a galaxy-galaxy-lensing cross-analysis is
\begin{equation}
    (\mathbf{C}_\ell)_{XY}=\left\{
    \begin{array}{ll}
      2(C_\ell^{\kappa\kappa}+N_\ell^{\kappa\kappa})^2/f_{\rm sky}^{\rm CMB} & X=Y=\kappa\kappa \\
      (C_\ell^{\kappa g_{a,i}}C_\ell^{\kappa g_{b,j}}+\delta_{ij}(C_\ell^{\kappa\kappa}+N_\ell^{\kappa\kappa})C_\ell^{g_{a,i} g_{b,i}})/f_{\rm sky}^{\cap} & X=\kappa g_{a,i}, Y=\kappa g_{b,j}\\
      \delta_{ij}(C_\ell^{g_{a,i} g_{c,i}}C_\ell^{g_{b,i} g_{d,i}}+C_\ell^{g_{a,i} g_{d,i}}C_\ell^{g_{b,i} g_{d,i}})/f_{\rm sky}^{\rm LSS} & X=g_{a,i}g_{b,i}, Y=g_{c,j}g_{d,j} \\
      2(C_\ell^{\kappa\kappa}+N_\ell^{\kappa\kappa})C_\ell^{\kappa g_{a,i}}/f_{\rm sky}^{\rm CMB} & X=\kappa\kappa, Y=\kappa g_{a,i} \\
      2C_\ell^{\kappa g_{a,i}}C_\ell^{\kappa g_{b,i}}f_{\rm sky}^{\cap}/(f_{\rm sky}^{\rm CMB}f_{\rm sky}^{\rm LSS}) & X=\kappa\kappa, Y=g_{a,i}g_{b,i} \\
      \delta_{ij}(C_\ell^{\kappa g_{b,i}}C_\ell^{g_{a,i} g_{c,i}}+C_\ell^{\kappa g_{c,i}}C_\ell^{g_{a,i} g_{b,i}})/f_{\rm sky}^{\rm LSS} & X=\kappa g_{a,i}, Y=g_{b,j}g_{c,j} \\
    \end{array} 
    \right. 
\end{equation}
and the Fisher matrix is
\begin{equation}
    F^{\rm 2D}_{ij}=\sum_{\ell=\ell_{\rm min}}^{\ell_{\rm max}}(2\ell+1)\sum_{XY} \frac{\partial C_\ell^{X}}{\partial \theta_i}(\mathbf{C}_\ell)^{-1}_{XY}\frac{\partial C_\ell^{Y}}{\partial \theta_j}
\end{equation}
where $f_{\rm sky}^{\cap}$ is the overlapping sky area between LSS and CMB surveys, and $X,Y$ sum over the available spectrum $\{\kappa\kappa, \kappa g_{a,i}, g_{a,i} g_{b,i}\}$ for all available galaxy tracers $a,b$ and redshift bins $i$ \cite{Sailer21}. For the purposes of our forecast we assume maximal overlap between LSS and CMB experiments, i.e.\ $f_{\rm sky}^\cap = \min(f_{\rm sky}^{\rm CMB},f_{\rm sky}^{\rm LSS})$. Note that the cross-spectra between tracers at different redshift bins is zero, so we do not sum over them. 

The simple treatment of priors is a strength of the Fisher formalism, with the inverse covariance of the priors simply treated as an added information matrix. We will be using conservative priors on the nuisance parameters to regularize forecast behavior. The set of priors used is shown in Table \ref{tab:prior}. Most of the nuisance parameter priors reflect the fidelity we have on constraining these parameters, by applying uncertainties proportional to their fiducial value. This cannot be done for parameters with a fiducial value of 0, so we describe this briefly. 
For $N_{ab,a\ne b}$ and $N_{2,ab,a\ne b}$, we apply the same errors as those of the auto-spectra, assuming there is full correlation between the shot noise of two galaxy populations (e.g.~$N_{ab,a\ne b} = 1/\sqrt{n_a n_b}$). The other terms with fiducial value have no prior, due to the lack of conservative expectation for their values. We have confirmed that our conservative priors have little effect on our final results. 
We will be considering both BBN and Planck priors for cosmological parameters.

\subsection{Benefits of multi-tracer cosmology}

The use of multiple tracers is closely associated with the idea of `sample variance cancellation' \cite{McDonald09,Bernstein11}.  For two quantities, e.g.~a density and velocity field, some relationships can be measured at the field level and thus do not pay any sample variance penalty.
For example, within linear theory and in general relativity, the density contrast $\delta(\mathbf{k})$ is related to the velocity divergence, $\theta(\mathbf{k})$ mode-by-mode.  A ratio of measured modes can thus measure $f$ with no sample variance penalty.  Similarly, for linear theory and scale-independent, deterministic bias the ratio of $\delta_i(\mathbf{k})$ for two galaxy samples measures the ratio of their biases in a way that is not limited by sample variance.  Colloquially we say that sample variance has `canceled', though it would be more correct to state that sample variance only enters when comparing a particular realization to the ensemble average.  While in general, cosmology dependence enters quantities (such as the linear power spectrum) that are expectation values, for some situations (such as the scale-dependent bias introduced by primordial non-Gaussianity) this is not the case and `cancellation' can lead to tremendous gains in precision.

Sample variance cancellation can be recast as a fitting problem, in which case it becomes the statement that when including cross-correlations in addition to auto-correlations noise that is highly correlated between probes does not get double counted \cite{White09}.  In such situations some eigenvectors `pay' the sample variance penalty while others do not.

Past applications of sample variance cancellation \cite{Blake13} have not found drastic gains in constraining power from easily achievable number densities and we are unlikely to have high-enough number densities for this either. However, another use of multiple tracers is to break degeneracies between parameters. Including multiple tracers allow us to distinguish between parameter dependencies of the data, breaking parameter degeneracies that limit our precision. 
Recent studies indicate that in a general case, it is this degeneracy breaking that drives the improvements in constraints, rather than sample variance cancellation \cite{Mergulhao22,Mergulhao23}.

\begin{table}[t]
    \centering
    \begin{tabular}{c|cc}
    Parameter & Definition & Fiducial value \\
    \hline
    $\log A_s$ & Primordial amplitude & $2.10732\times10^{-9}$\\
    $n_s$ & Spectral index & 0.96824\\
    $h$ & Hubble parameter: $H_0/$100 km/s/Mpc & 0.6770 \\
    $\tau$ & Optical depth to reionization & 0.0568 \\
    $\omega_{c}$ & fractional dark matter density: $\Omega_{c}h^2$ & 0.11923 \\
    $\omega_b$ & fractional baryonic matter density: $\Omega_b h^2$ & 0.02247 \\
    \hline
    $b$ & First order bias $\delta_g\supset b_g\delta_m$ & $b_{\rm LAE}$: Eq. \ref{bias_fit}\\
    & & $b_{\rm LBG}$: Eq. \ref{b(m,z)} \cite{Wilson19}\\
    $b_2$ & Second order bias &  Eq. \ref{sheth-tormen}\\
    $b_s$ & Shear bias & $-\frac{2}{7}(b_g-1)$\\
    $\alpha_0$ & Lowest order counter term $P_{aa}\supset\alpha_{0,a}k^2 P_m$& $1.22+0.24 b_a^2 (z-5.96)$\\
    $\alpha_{2n}$ & Higher order counter terms $P_{aa}\supset\alpha_{2n,a}\mu^{2n}k^{2} P_m$& $0$\\  
    $N_{ab}$ & Shot noise term $P_{ab}\supset N_{ab}$ & $N_{aa}=1/\Bar{n_a}$\\
    & & $N_{ab,a\ne b}=0$ \\
    $N_{2,ab}$ & FOG-like effect $P_{ab}\supset N_{2,ab}(k\mu)^2$ & $-N_{ab}\sigma_v^2$ \\
    $N_{4,ab}$ & FOG-like effect $P_{ab}\supset N_{4,ab}(k\mu)^4$ & 0\\
    \end{tabular}
    \caption{The parameters for our fiducial cosmological model, $\Lambda$CDM with massive neutrinos. The cosmological parameters are placed on the top and model parameters are on the bottom. The model parameters are individually assigned either to each tracer or each spectra. Furthermore, we allow model parameters to evolve over redshift, which gives us new degrees of freedom for each additional $z$ bin.
    }
    \label{tab:fiducial}
\end{table}

\begin{table}[]
    \centering
    \begin{tabular}{c|c||c|c||c|c}
        Parameter & Uncertainty & Parameter & Uncertainty & Parameter & Uncertainty \\
        \hline
        $b$ & 100\% & $N_{aa}$ & 100\% & $\alpha_0$ & 100\% \\
        $b_2$ & 100\% & $N_{ab,a\ne b}$ & $\sqrt{N_{aa} N_{bb}}$ & $\alpha_2$ & $\infty$ \\
        $b_s$ & 100\% & $N_{2,aa}$ & 300\% & $\alpha_4$ & $\infty$ \\
        $N_{4}$ & $\infty$ & $N_{2,ab,a\ne b}$ & $3\sqrt{N_{2,aa}N_{2,bb}}$               
    \end{tabular}
    \caption{Prior uncertainties for nuisance parameters. For $\Lambda$CDM parameters, we use BBN or CMB primary priors appropriately.}
    \label{tab:prior}
\end{table}

\section{Multitracer} \label{multitracer}

Before we turn to forecasts for particular models let us investigate some test samples with semi-realistic parameters to gain a better understanding of the multitracer approach. To do this, we make forecasts for the amplitude of the linear theory power spectrum, $\sigma_8$, holding fixed the other $\Lambda$CDM parameters $\{h,\tau,n_s,\omega_c,\omega_b \}$ at some fiducial values.  We allow the biases, counterterms and stochastic terms for each sample to vary.

Throughout this section we use the notation $a$ and $b$ to refer to the auto-spectrum analyses of each tracer, ``auto'' or $aa+bb$ refers to using two tracer populations, but with only the galaxy auto-spectra considered (i.e.~incomplete multitracer) and ``full'' or $a\times b$ refers to the complete multitracer approach including the galaxy-galaxy cross-spectrum. 

\subsection{Linear}

We will start by investigating multitracer cosmology in the Fisher formalism using linear theory. In linear theory, with the exception of shot noise, the power spectra of each galaxy population is proportional to the linear matter power spectrum
\begin{equation}
    P_{ab}(k,\mu) = (b_a+f\mu^2)(b_b+f\mu^2) P_L(k) + N_{ab}
\end{equation}
where $b$ is the (deterministic, scale-independent) linear bias, $f$ is the scale-independent growth factor, $P_L$ is the linear theory matter power spectrum, $N_{ab}$ is the shot noise term, and $\mu$ is the cosine of the angle to the line of sight.
The relevant derivatives for linear theory power spectrum are expressed as
\begin{equation}
    \begin{split}
        \frac{\partial P_{ab}}{\partial \log\sigma_8}&=2(b_a+f\mu^2)(b_b+f\mu^2)P_L\\
        \frac{\partial P_{ab}}{\partial b_c}&=[\delta_{ac}(b_b+f\mu^2) +\delta_{bc}(b_a+f\mu^2)] P_L\\
        \frac{\partial P_{ab}}{\partial N_{cd}}&=\delta_{ab,cd}
    \end{split}
\end{equation}
We use this method first as a check on our code (since the calculation can be done analytically) and second to illustrate some of the features of multitracer cosmology. 

The Fisher information is a function of both the amount of data and the degeneracy between the derivatives of the data with respect to the parameters. 
In multitracer cosmology, we hope to see an increase in constraints that goes beyond the simple increase in data quantity, which must come from breaking of degeneracy. However, in linear theory, the lack of parameter dependencies beyond the $\sigma_8$-$b$ degeneracy suggests that this will not occur. 

At the minimum, we expect the low-bias tracer to improve the constraints by a significant margin, as low-bias tracers have an inherent advantage in constraining structure growth compared to high-bias tracers. Even if there is no synergy between the measurement, matching the low-bias constraints will significantly reduce the error from having a single, high-bias tracer. This can easily be explained in the following sense. 
In linear theory, the auto-spectrum of a biased tracer is given by 
\begin{equation}
    P(k,\mu)=(b+f\mu^2)^2P_L(k)+N=(b\sigma_8+f\sigma_8\mu^2)^2\hat{P_L}(k)+N
\end{equation}
where $\hat{P_L}$ is the normalized linear power-spectrum, $P_L=\sigma_8^2\hat{P_L}$. As $b$ is a nuisance parameter, one must probe for the $\mu$-dependence of the spectrum to obtain $\sigma_8$. With the error proportional to the overall spectrum mode-by-mode ($\delta P\propto P$), one loses constraining power on $\sigma_8$ with greater $b$.
To maximize SNR, it's clear that we must minimize $b$ and maximize $\Bar{n}$, both of which are satisfied by choosing a low-bias sample. 
In the limit that we go to $b=0$, we are measuring the velocity divergence $\theta$. In this case, our precision of the amplitude of $P$ is sample variance limited such that $\delta A/A=\sqrt{2/N_{\rm modes}}$, making $\delta \sigma_8/\sigma_8=(1/2)\sqrt{2/N_{\rm modes}}$. 

Before proceeding further, let us define an effective shot noise $\Tilde{N}$, as
\begin{equation}
    b^2 \Tilde{N} \equiv \frac{1}{\Bar{n}}
\end{equation}
We will use $\Tilde{N}$ to explore different noise regimes throughout this section. For reference, at $z=3$, LAEs with $f_{\rm lim}$ of 3, 5, 10$\times 10^{-17}\,\text{erg}\,\text{cm}^{-2}\,\text{s}^{-1}$ have $\Tilde{N}\approx$ 100, 170, 410, and LBGs with $m=$ 24, 24.5, 25 have $\Tilde{N}\approx$ 570, 230, 140, respectively, in units of $h^3\,\text{Mpc}^{-3}$. 

\begin{figure}
    \centering
    \includegraphics[width=\textwidth]{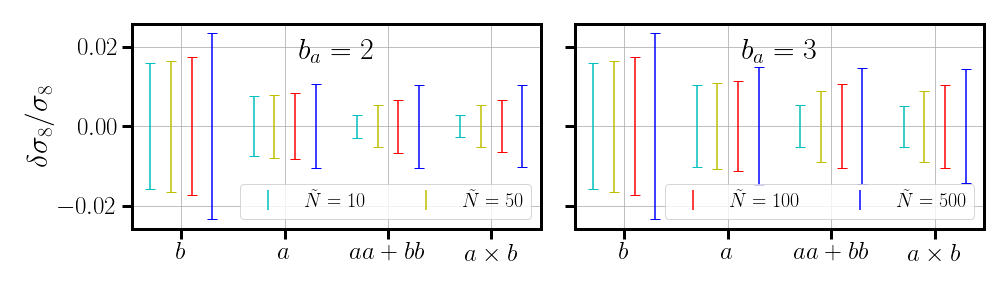}
    \caption{The relative constraints of $\sigma_8$ computed in linear theory, with constant linear biases and number densities. $\Tilde{N}=(b^2 \Bar{n})^{-1}$ is the effective shot noise. $a$ and $b$ refer to the one-tracer constraints from each tracer, $aa+bb$ refer to the auto-spectra (incomplete multitracer) constraints, and $a\times b$ refer to the complete multitracer constraint.
    }
    \label{fig:l_ds8}
\end{figure}

\begin{figure}
    \centering
    \includegraphics[width=\textwidth]{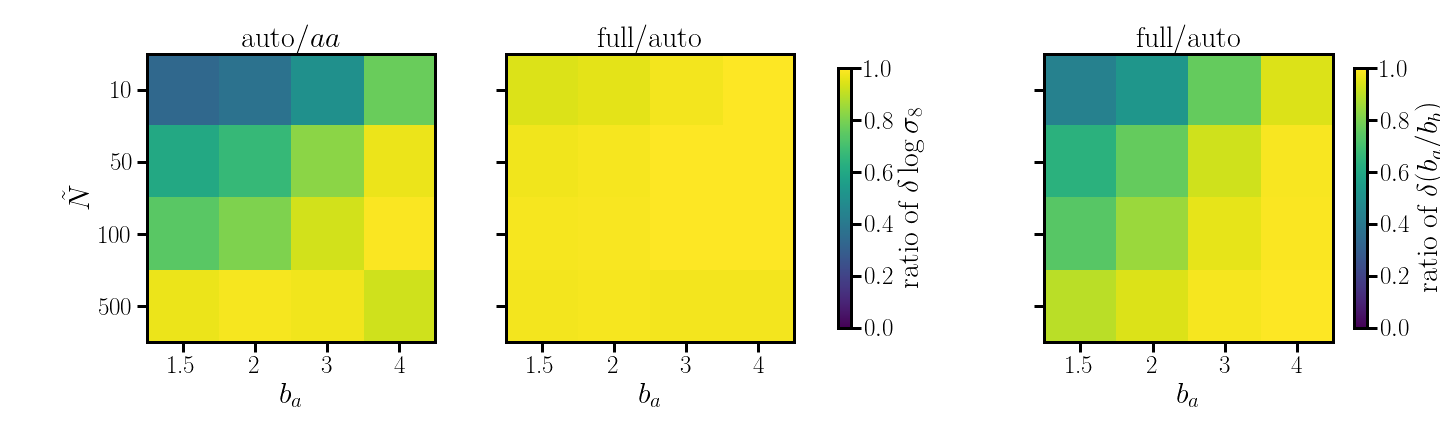}
    \caption{Left: Showing improvement in relative constraints of $\sigma_8$, when including the high bias tracer to the low bias (auto/$aa$) and including cross-spectrum to auto-spectra (full/auto). Right: Showing improvement in relative constraints of $b_a/b_b$ when including cross-spectrum to auto-spectra (full/auto). The significant improvement in $\delta(b_a/b_b)$ compared to $\delta\log\sigma_8$ show the power of sample variance cancellation where present.}
    \label{fig:ratios_linear}
\end{figure}

Let us consider samples in linear theory, with constant biases, $b_a=\{1.5,2,3,4\}$, $b_b=5$, and a linear regime $k_{\rm max}=0.1\,h\,\text{Mpc}^{-1}$. A subset of our constraint calculations are shown in Figure \ref{fig:l_ds8}. 
As we predict there is little to no improvement from adding the cross-spectrum, as shown in Figure \ref{fig:ratios_linear}, indicating that we are losing little to degeneracy.  

In the same figure, we detect that the `auto' ($aa+bb$) results are significantly better than the single low-bias result. This indicates that a naive error propagation of $1/\sigma^2=1/\sigma_1^2+1/\sigma_2^2$ is not appropriate, as this predicts the low-bias and `auto' constraints to be similar for most of our samples.  The reason is that we must include the correlation between measurements (e.g.\ the velocity part of the power spectra is the same).  Including the correlation, $r$, the variance $\sigma^2$ becomes
\begin{equation}
    \sigma^2 = \frac{(1-r^2)\sigma_1^2 \sigma_2^2}{\sigma_1^2-2r\sigma_1\sigma_2+\sigma_2^2}
    \label{variance_linear}
\end{equation}
for the composite measurement, with $\sigma_i$ referring to the uncertainty of the $i$'th measurement. Refer to Appendix \ref{variance_calc} for detailed calculations.

This result explains the two main trends that we see. The first trend, the decreasing improvement as one takes $b_a\rightarrow b_b$, can be illustrated by the case with redundant measurements ($\sigma_1=\sigma_2$, $r\rightarrow1$), which yields $\sigma^2=\sigma_1^2=\sigma_2^2$. The second trend, the increasing improvement as one goes to lower noise, follows the limit of non-redundant measurements with high correlation ($\sigma_1\ne\sigma_2$, $\abs{r}\rightarrow1$), which yields $\sigma^2\rightarrow0$.

One can directly compare the $\delta\sigma_8/\sigma_8$ results with improvements in constraints for the linear bias ratio $b_a/b_b$, where sample variance cancellation is present \cite{McDonald09}. A direct comparison is made in Figure \ref{fig:ratios_linear}. The lack of improvement in $\delta\log\sigma_8$ by including the cross-correlation, as compared to the significant improvement in $\delta(b_a/b_b)$, illustrates that sample variance cancellation is not relevant for constraining $\sigma_8$. 

\subsection{Nonlinear}

Now let us study the nonlinear scenario, with the fully general power spectrum (Eq.~\ref{general_spectra}). The existence of many more parameters allows for the existence of significant parameter degeneracies, and hence the potential for significant degeneracy breaking from multitracer techniques.  We test on samples of identical linear biases and noise to the linear regime examples, but with further analysis including CMB lensing as well.

\begin{figure}
    \centering
    \includegraphics[width=\textwidth]{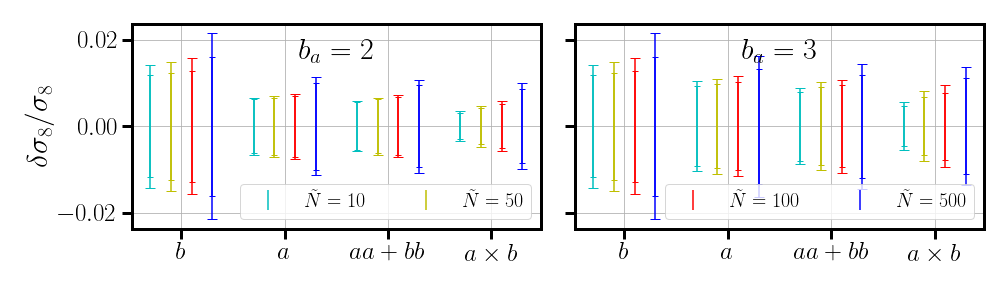}
    \caption{The relative constraints of $\sigma_8$ in the nonlinear regime, with various noise and $b_a$ values. The constraints with smaller caps show the respective constraints including lensing.
    }
    \label{fig:nl_ds8}
\end{figure}

A subset of the results are shown in Figure \ref{fig:nl_ds8}. 
For some samples there is visible improvement of constraints from adding the $ab$ cross-spectrum, contrary to the marginal improvement seen in the linear regime. This indicates that we have degeneracy breaking at hand, as the test case in the linear regime has shown that the increase in information from the additional spectrum only improves our constraints marginally. 
The average improvement in constraining power by adding the cross-spectrum (auto$\to$full), which can be interpreted as the average degeneracy breaking, is 17\%. This marks a 22\% improvement from the low-bias tracer $a$ and a 68\% improvement from the high-bias tracer $b$.
This effect of degeneracy breaking can vary significantly with the noise regime that we are working in.  For instance, at $\Tilde{N}=10$ the improvement from auto$\to$full is 30\% on average, while at $\tilde{N}=500$ it is only 5\%. 

One can hope to further improve these results by reducing the number of added free parameters. A common practice in the field is to neglect the cross-spectrum stochastic terms $N_{2n,ab}$, by assuming that the two galaxy tracers have completely disjoint inhabitance of halos. While past work has shown that the inclusion of these parameters do not crucially change the forecast behavior \cite{Mergulhao23}, it will be worthwhile to investigate this further by considering small-scale cross-clustering of LBGs and LAEs at the redshifts and luminosities of relevance to future cosmology surveys. 

From the perspective of added parameter space, the lack of new parameters other than $\alpha_{0,\kappa}$ makes CMB lensing an ideal candidate for cross-analysis.
Our results, also shown in Figure \ref{fig:nl_ds8}, indicate that we have 11\% improvement on average when adding lensing to our full multitracer case. When adding to each one-tracer case, we see a 6\% improvement and 19\% improvement for tracers $a$ and $b$, respectively. Although the increments are significantly less than that of the galaxy-galaxy multitracer analysis, it is worthwhile to note that there is still improvement from lensing after including two galaxy tracers.
We may attribute the little improvement from lensing to the limitations placed on the number of modes measured in a 2D power-spectrum, which is significantly less than that of the 3D spectrum, and the lack of cosmological parameters in this ``fixed-shape'' approach. 

\section{Radiative transfer}
\label{sec:RT}

Radiative transfer (RT) is a known, yet unresolved potential cause of concern for the use of LAEs as matter density tracers. Ref.~\cite{Zheng11} suggests that RT can significantly vary the number of galaxies in catalog depending on the local density and LOS velocity divergence. The latter effect is degenerate with the $\mu$ dependence used to constrain $f\sigma_8$ in RSD studies (refer to \S\ref{multitracer} for a brief discussion in linear theory). A more recent work \cite{Behrens18} argues that the limited resolution of early simulations are the source of such a large effect. Simulations with higher resolution show that regions of Ly$\alpha$ emission tend to have higher density, such that the photons escaping the galaxy have shifted further away from resonance and thus are less susceptible to scattering by the surrounding environment. While direct observations of RT effect are currently unavailable, the impact of H{\sc i} absorption on LAE clustering has been detected \cite{Momose21}.

For an analysis using only LAEs, a large RT contribution to the selection of galaxies would lead to very large errors on $\sigma_8$. However, with a complex model that can potentially break degeneracies through multi-tracers, these samples may prove to be useful. In order to make a preliminary investigation of whether RT-affected tracers can still be valuable in multi-tracers, we take the following approach. We imagine that the RT effect is small enough that it can be included by taking
\begin{equation}
    P(k,\mu) \to P(k,\mu) + 2\delta f\left( b\mu^2+f\mu^4 \right) P_L + \mathcal{O}(\delta f^2)
\end{equation}
while leaving the higher loop terms unchanged.  In principle there are corrections to all of the higher order biases and the 1-loop terms as well, but we are assuming that these terms are sub-dominant and multiplied by a ``small'' RT effect and thus they can be neglected.  A more complete and consistent calculation will be attempted in a future publication.

\section{Results} \label{sec:results}

In this section we outline our main forecast results for two experiment designs: `Stage 4.5', motivated by DESI II, and `Stage V'. As discussed in \S\ref{tracers} we will use one luminosity limit ($f_{\rm lim}=10\times10^{-17}$ for LAEs and $m=24.5$ for LBGs) as the fiducial design for Stage 4.5 and consider multiple scenarios of Stage V luminosity limits for a preliminary optimization constrained on the total number of galaxies observed. The experimental designs are shown in Table \ref{tab:experiments}. 

In this section, we will consider gains from any additional tracer as ``multi-tracer improvement'', e.g.~improvements from LBG-only analysis to an LAE$\times$LBG or $\kappa\times$LBG analysis. This is unlike \S\ref{multitracer} where the focus was on the improvement gained through the inclusion of cross-spectra, and hence now the appropriate cross-spectra are always included when multiple tracers are in use. 
It should not be a surprise, however, that an inclusion of additional tracer improves the constraints with all else unchanged. To distinguish the gains from multitracer from the simple increment in data, we will be adjusting the number density of the LBG data available when including LAEs. The constraint here will be the total number of galaxies observed, as we have done when deciding the optimization for Stage V. As we have shown in Figure \ref{fig:show_spectra} that LBGs are less shot-noise dominated with existing luminosity cuts and the number density of LAEs are several factors smaller than LBGs, we will observe LAEs at maximum efficiency and decrease the LBG observation efficiency necessarily, to 0.61 (note that these efficiencies are multiplied by the 50\% success rate for redshift detection that is always in place)\footnote{We have confirmed that the lack of inclusion of this LBG efficiency decrease affects the improvement results by $<10\%$}. A similar treatment cannot be made for lensing as they are provided by CMB surveys separate from LSS experiments. Hence, we will compare the galaxy-only and $\kappa\times$galaxy with the same galaxy samples and for this reason the LAE and lensing multitracer improvements should not be regarded on the same basis. 

\begin{figure}
    \centering
    \includegraphics[width=\textwidth]{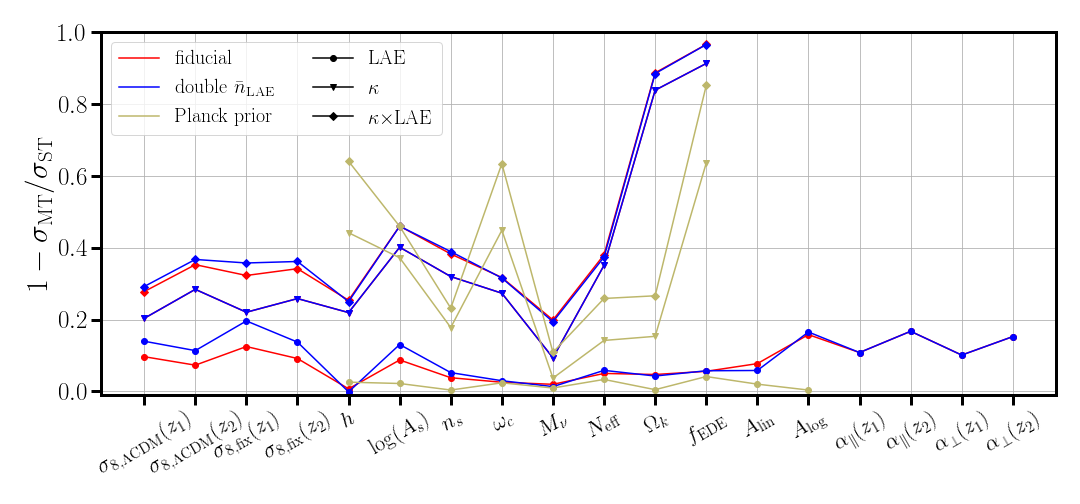}
    \caption{A summary of parameter constraint improvements from using multitracers in the Stage 4.5 design. The `ST' and `MT' on the y-axis refer to single tracer and multitracers, where the LBG-only sample is considered the base single tracer sample. The type of marker depicts what tracer was added on to the sample. The shown improvements for $f_{\rm EDE}$, $A_{\rm lin}$, $A_{\rm log}$ are the average improvements across the parameter space probed.}
    \label{fig:summary}
\end{figure}

For all forecasts using full shape information, we include all relevant $\Lambda$CDM and nuisance parameters (Table \ref{tab:fiducial}) and treat extra parameters of interest as one parameter extensions. For cosmological priors, we will use conservative BBN priors of $\sigma(\omega_b)=0.0005$ or CMB primary priors from Planck. A summary of the multi-tracers improvement in the Stage 4.5 design are provided in Figure \ref{fig:summary}. 

The blue curve in the figure shows a case for when LAE number densities are doubled, in the BBN prior scenario. This is to address the concern raised in \S\ref{LAEs} that offsets of factor 2 on LAE number density estimates are not unlikely. The results illustrate that there are no qualitative changes to the final results. 

\subsection{Growth rate} \label{growth rate}

In this section, we measure structure growth by measuring $\sigma_8$ constraints. While $\sigma_8$ is a well-motivated variable to pursue in multi-tracers both in terms of existing tensions in the field and linear theory arguments provided in \S\ref{multitracer}, it is also a variable where we expect significant impact from RT effects as mentioned in \S\ref{sec:RT}. It is possible that multi-tracers break the RT-induced degeneracies enough such that inclusion of LAEs significantly improve $\sigma_8$ constraints. 

We use two approaches in this section. First we have our full-shape approach, which marginalizes over all of the $\Lambda$CDM parameters. Another approach is to use the `fixed-shape' approach of \S\ref{multitracer}, which takes a model-agonistic approach and only marginalizes over amplitude. The latter approach marginalizes over a subset of the former approach's parameters and thus always yields tighter constraints. 
In both approaches, the lensing auto-spectrum is not included. This ensures that the $\sigma_8(z)$ constraints are local, using only the data collected in each redshift bin up to the shared BBN prior. 

\begin{figure}
    \centering
    \includegraphics[width=\textwidth]{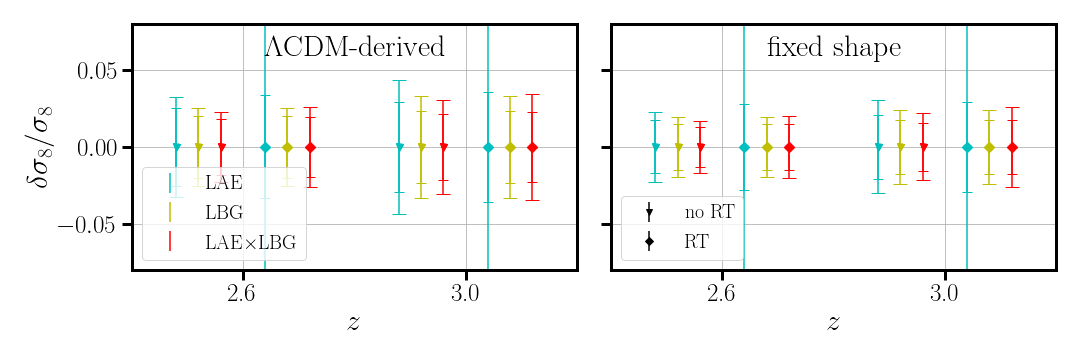}
    \caption{Relative constraints on $\sigma_8(z)$ from Stage 4.5 in the $\Lambda$CDM-derived (left) and fixed shape (right) methods. The smaller cap sizes correspond to the constraints including lensing. We see that the inclusion of RT effects diminish majority of the improvement from including LAEs.}
    \label{fig:result_D2_s8}
\end{figure}

\begin{figure}
    \centering
    \includegraphics[width=1.\textwidth]{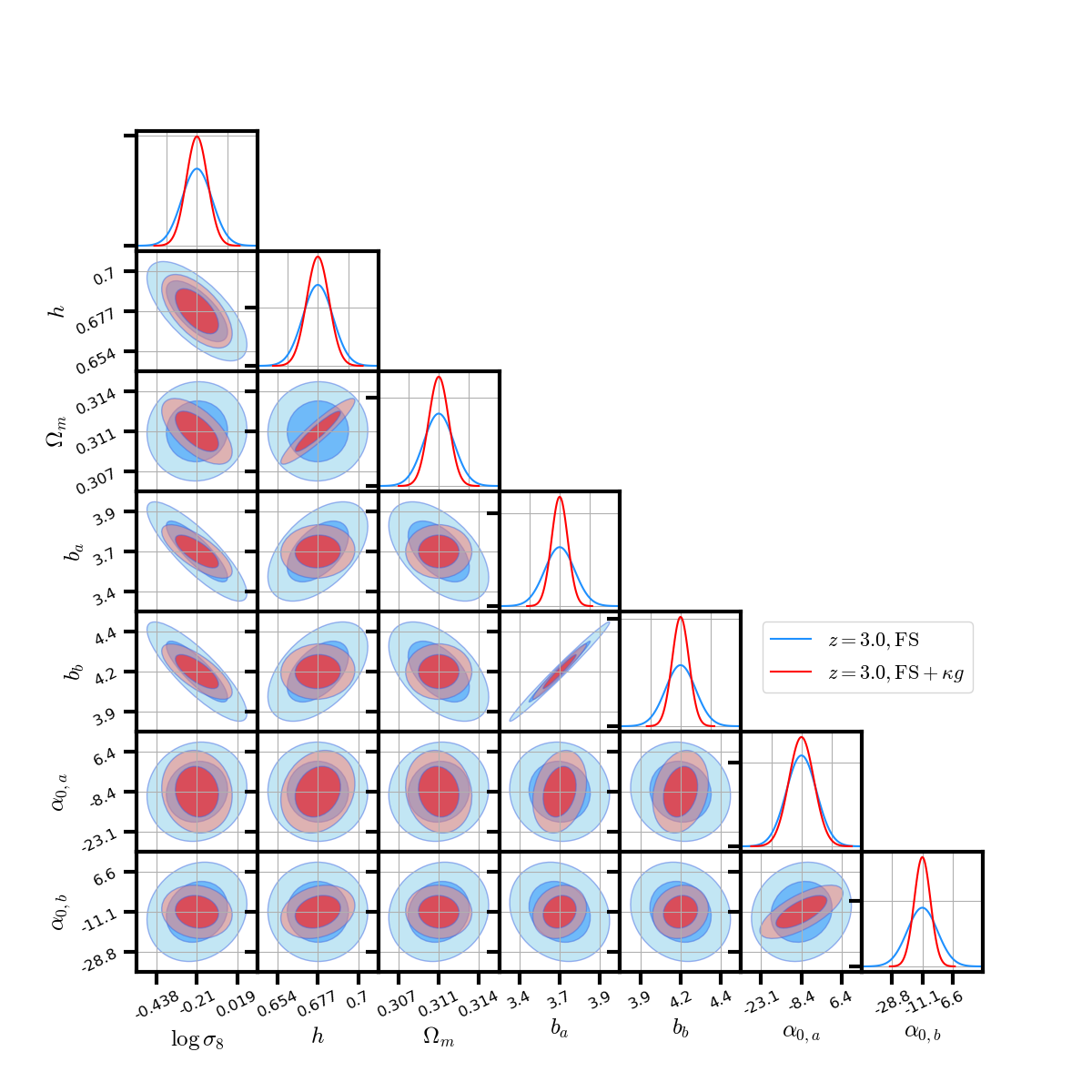}
    \caption{The marginalized contours of the full-shape approach to constraining growth. The contours shown are from the $z=3$ bin of Stage 4.5, with and without galaxy-lensing cross-spectra. The contours show clear signs of degeneracy breaking from the addition of lensing. In particular, the correlation coefficient between cosmological parameters $\Omega_m$ and $h$, and linear biases $b$ are significantly reduced from $\sim0.5$ to $\lesssim0.05$. Note that degeneracy breaking does not necessarily reduce of correlation coefficients.}
    \label{fig:contour_rotate}
\end{figure}

The results for Stage 4.5 are shown in Figure \ref{fig:result_D2_s8}.
We find that the both approaches gain $\approx 10\%$ from the introduction of LAEs. This is less than the improvements from including lensing, which are $\approx20-30\%$. These improvements in total amount to $\approx25-35\%$. Notice that unlike in \S\ref{multitracer}, the low bias tracer LAE do not have tighter $\sigma_8$ constraints than the high bias tracer LBG. This is due to LAEs being more shot noise dominated than LBGs at the luminosity limits of Stage 4.5. This increases the importance of lensing as well. As discussed in \S\ref{multitracer}, the primary contributions to constraint improvement come from degeneracy breaking. This is visualized in Figure \ref{fig:contour_rotate}, which shows the marginalized contours of the full-shape approach at the $z=3$ bin of Stage 4.5. In particular, the correlation coefficient between cosmological parameters $\Omega_m$ and $h$, and linear biases $b$ are significantly reduced from $\sim0.5$ to $\lesssim0.05$ by the addition of galaxy-lensing cross-spectra. This indicates that the cross-spectra has broken degeneracy between linear bias and cosmology, as we have predicted. Note that degeneracy breaking does not necessarily reduce correlation coefficients and thus this only highlights some of the most noticeable degeneracy breakings. 
When RT is present, multi-tracer benefits from LAEs diminishes significantly with the constraints weakening by approximately 5\%. This turns to an improvement when lensing is included ($\kappa\times$LBG$\to\kappa\times$LAE$\times$LBG), but the improvements remain less than 5\%. 

\begin{figure}
    \centering
    \includegraphics[width=\textwidth]{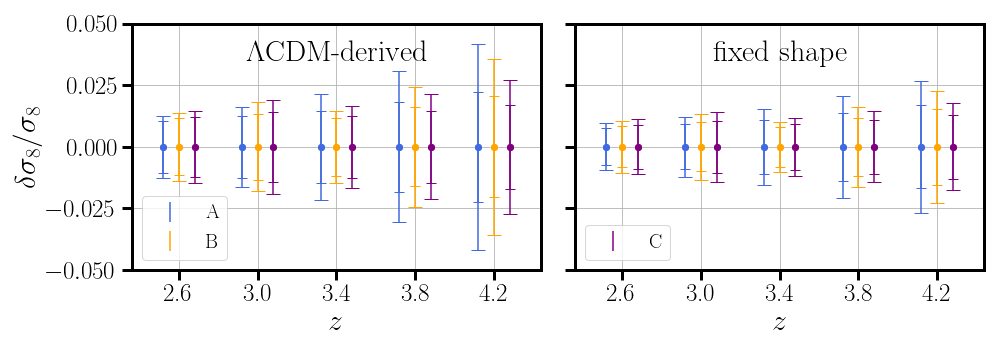}
    \caption{Relative constraints on $\sigma_8(z)$ from Stage V surveys in the $\Lambda$CDM-derived (left) and fixed shape (right) methods, with both LAEs and LBGs, without RT. The smaller cap sizes correspond to the constraints including CMB lensing. `A', `B', and `C' each correspond to Plans A, B, and C of Stage V (Table \ref{tab:experiments}). All error bars assume inclusion of all multi-tracer spectra.}
    \label{fig:SV_s8}
\end{figure}

Figure \ref{fig:SV_s8} shows the Stage V constraints for all 3 scenarios, with both galaxy tracers included without RT. While there are sizable differences between scenarios, they are not drastic, especially with the inclusion of lensing. This indicates that $\sigma_8$ constraints are robust to experiment designs. 

\subsection{Standard model}

\begin{figure}
    \centering
    \includegraphics[width=\textwidth]{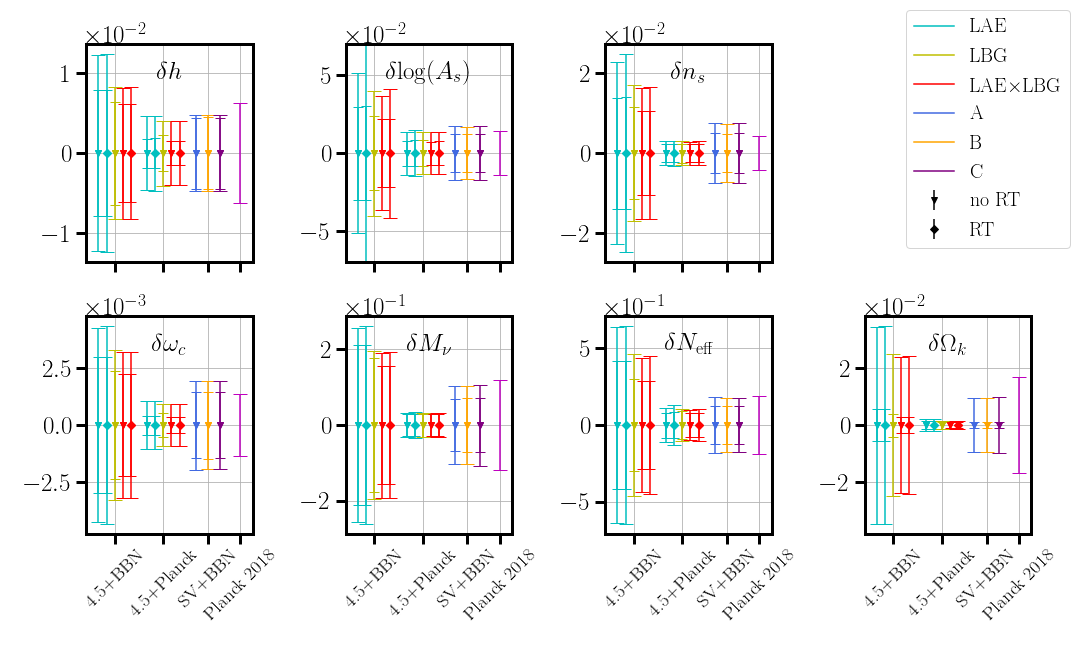}
    \caption{$\Lambda$CDM and extension results. The `4.5+BBN', `4.5+Planck' and `SV+BBN' each refer to Stage 4.5 with a BBN prior, Stage 4.5 with a Planck prior, and Stage V with a BBN prior, respectively. `A', `B', and `C' each correspond to Plans A, B, and C of Stage V (Table \ref{tab:experiments}). The massive neutrino constraints include DESI BAO and lensing auto-spectra for all results. The magenta errorbars are the constraints given by Planck \cite{PCP18}.
    }
    \label{fig:LCDM_results}
\end{figure}

Multi-tracers will further assist next-generation LSS experiments to provide constraints similar to CMB experiments alone. 
Figure \ref{fig:LCDM_results} shows a summary of forecasts. It does not include $\omega_b$ as it shows no influence from either multi-tracers nor RT, likely due to the prior dominating the constraining power.
Forecasts indicate that with a BBN prior, multi-tracers through LAEs will improve the constraints on all other cosmological parameters by $\lesssim10\%$. Inclusion of lensing generally yields larger improvements, with the improvement varying from $\sim20\%$ for $h$ to $\sim85\%$ for $\Omega_k$. Overall, the improvements amount to $25-45\%$, with the exception of $\Omega_k$ at 89\%. 

With Planck primary priors, there are varying multi-tracer effects. Multi-tracers with LAEs on a LBG-only sample will have even less improvement compared to when using BBN priors, at $\sim5\%$, but when lensing is included (i.e.~LBG$\times\kappa\to\text{LAE}\times\text{LBG}\times\kappa$) the improvements increase, marking $\sim35\%$ for $h$ and $\omega_c$. Similarly, the improvements from adding lensing vary across parameters, marking a wide range between $\sim20\%$ to $\sim65\%$. $h$ and $\omega_c$ are again most affected, with constraints improving $\sim60\%$ when adding lensing to the LAE$\times$LBG sample. 
Refer to Figure \ref{fig:summary} for specific parameter constraint improvements. 

Inclusion of RT generally has little effect on the cosmological constraints, except for $\log(A_s)$ that experience significant effects for forecasts without $\kappa$ due to reasons similar to $\sigma_8$. 
The final constraints with inclusion of all tracers experience sub-percent effect from RT, however. 
Our results also indicate that Stage V designs provide constraints significantly stronger than that of Stage 4.5, regardless of the luminosity limits adopted. 

\subsection{Distances}

One of the most important constraints from LSS surveys are the distance measurements made through the BAO peaks in the power-spectra. Previous work have forecasted that near-term surveys such as DESI, Euclid and HIRAX can measure angular diameter distance and Hubble parameter to sub-percent and percent precision, respectively, up to $z\sim2$, and future experiments to sub-percent precision up to $z\sim5$ \cite{Sailer21}. 
We will follow the modelling methods used in ref.~\cite{Sailer21}, marginalizing over linear bias and 15 polynomial coefficients defined as 
\begin{equation}
    P_{\rm obs} = \alpha_\parallel^{-1}\alpha_\perp^{-2}P_{\rm recon} + \sum_{n=0}^4 \sum_{m=0}^2 c_{nm}k^n \mu^{2m}
\end{equation}
where $\alpha_\parallel$ and $\alpha_\perp$ are A-P parameters and $P_{\rm recon}$ is the post-reconstruction power-spectrum calculated within the Zeldovich approximation using the ``RecSym'' convention \cite{White15b,Chen19b}. We refer the reader to ref.~\cite{Sailer21} for further details about the model.

\begin{figure}
    \centering
    \includegraphics[width=\textwidth]{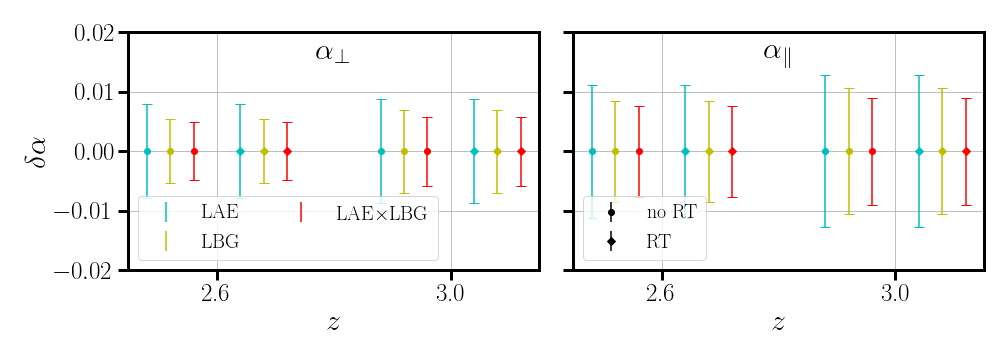}
    \caption{Stage 4.5 constraints on A-P parameters ($\alpha_\perp$, $\alpha_\parallel$), which can be interpreted as relative constraints on BAO parameters ($D_A(z)/r_d$, $r_d H(z)$), respectively. The inclusion of RT clearly have minimal effect for BAO constraints.}
    \label{fig:BAO}
\end{figure}

Our results, shown in Figure \ref{fig:BAO}, show that Stage 4.5 will reach sub-percent precision in both distance measures in its entire redshift range $2.4<z<3.2$, with the aid of multi-tracers improving constraints by $\sim10-15\%$. 
More significantly, these improvements are uninfluenced by the presence of RT. This is not surprising, given that BAOs are a well-localized feature with a known shape and the modelling process already marginalizes over any $\textbf{k}$ dependence for each spectrum beyond $P_{\rm recon}$.

\begin{figure}
    \centering
    \includegraphics[width=\textwidth]{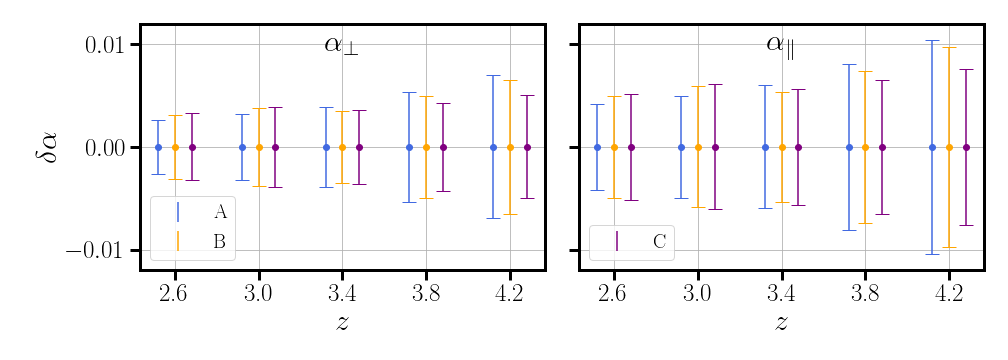}
    \caption{Stage V constraints for A-P parameters ($\alpha_\perp$, $\alpha_\parallel$), which can be interpreted as relative constraints on $D_A(z)/r_d$ and $r_d H(z)$, respectively. `A', `B', and `C' each correspond to Plans A, B, and C of Stage V (Table \ref{tab:experiments}). Each constraint assumes the inclusion of all multi-tracer spectra.}
    \label{fig:SV BAO}
\end{figure}

Forecasts for Stage V surveys, shown in Figure \ref{fig:SV BAO}, indicate that future survey of such kind can improve and further extend these constraints up to $z=4.4$. 
The comparable results between the different designs of Stage V indicate that such constraints are robust to the survey optimization, although fractional improvements can be gained at some redshifts in price for constraining power at other redshifts.

\subsection{Neutrino mass and light relics}

Variation of neutrino mass is a reasonable extension to the $\Lambda$CDM model. Measurements have yet to determine whether the neutrino mass hierarchy is normal ($m_{\nu_2}<m_{\nu_3}$, $\Sigma m_{\nu_i}\gtrsim 58$meV) or inverted ($m_{\nu_2}>m_{\nu_3}$, $\Sigma m_{\nu_i}\gtrsim 105$meV) \cite{CMBS4}. 
Sum of neutrino masses can be constrained cosmologically by measuring its effect on power-spectra. As neutrinos transition from relativistic to non-relativistic species at scale factor $a_{\rm nr}$, neutrinos only cluster at large scales ($k\ll k_{\rm nr}=a_{\rm nr}H(a_{\rm nr})$), and damp the power-spectra on smaller scales \cite{Sailer21}. 

For neutrino mass ($M_\nu=\Sigma m_{\nu_i}$) forecasts, we include DESI BAO and SO lensing auto-spectrum for all results. The DESI BAO data included is from the Bright Galaxy Survey (BGS; \cite{Hahn23,DESI}), Luminous Red Galaxies (LRG; \cite{Zhou23,DESI}), and Emission Line Galaxies (ELG; \cite{Mostek13}), post-reconstruction. We do not include the Lyman-alpha data due to its redshift overlap with next-generation facility galaxies.
The results shown in Figure \ref{fig:LCDM_results}, show that Stage 4.5 will be able to constrain the total neutrino mass to $157$meV ($28$meV) with BBN (Planck) priors. 
While improvements from individual tracers are $\lesssim10\%$ for BBN priors and $\lesssim4\%$ for Planck priors, the total improvement from both LAEs and lensing combined are double that, at 20\% and 11\% for BBN and Planck priors, respectively. 

\begin{figure}
    \centering
    \includegraphics[width=1.\textwidth]{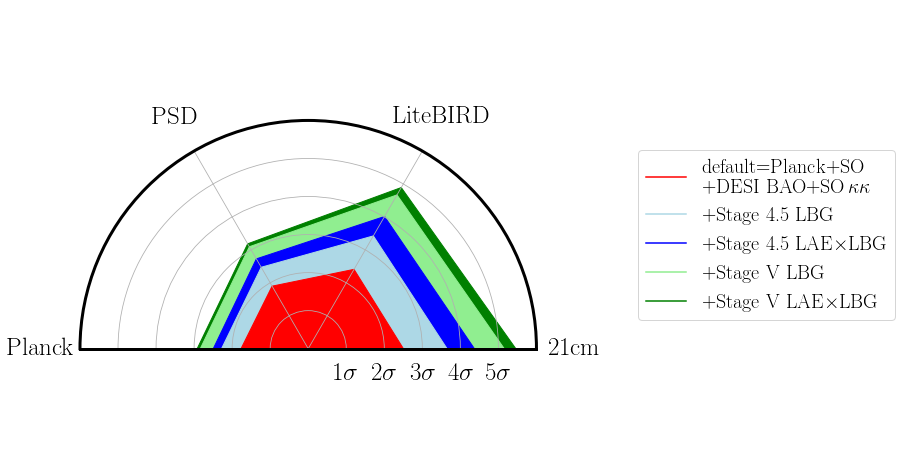}
    \caption{Fidelity of detection for normal neutrino mass hierarchy given $\tau$ constraints provided by external facilities. Each contour represents different information used for forecast. The red contour is only the default set of data, from Planck+SO CMB prior, DESI BAO measurements, and SO lensing auto-spectrum. The other contours include Stage 4.5 of Stage V galaxies and appropriate galaxy-lensing cross-correlations in addition to the default set. The labels show external constraints on $\tau$ that can be obtained currently or in the near-future. $\sigma(\tau)$ in each case are 0.0071 \cite{PlanckLegacy18,Sailer21,Pagano20}, 0.0058 \cite{Sailer21,Sailer22}, 0.002 \cite{Allys22,Valentino18}, and 0.0009 \cite{Sailer22}, for Planck, Planck+SO+DESI (PSD), LiteBIRD, and 21cm surveys, respectively.
    In particular, the constraint provided by LiteBIRD is the cosmic variance limit \cite{Watts18} for CMB experiments. 
    }
    \label{fig:Mnu_tau}
\end{figure}

Ref.~\cite{Liu:2015txa,Chudaykin19,Sailer21,Sailer22} notes that improvement of $\tau$ measurements will significantly tighten $M_\nu$ measurements by removing the $\tau$-$A_s$ degeneracy in primary CMB. Ref.~\cite{Sailer24} has shown that this leads to immense advantages when considering the $\tau$-constraints provided by near-future CMB surveys. Multiple galaxy tracers allow us to further take advantage of this. Figure \ref{fig:Mnu_tau} shows the effect, by showing forecasts for $\sigma$'s towards normal hierarchy detection using external $\tau$ constraints from current and near-future sources. In particular, LiteBIRD \cite{LiteBIRD} will be able to provide cosmic variance limited constraints on $\tau$ for CMB experiments at $\sigma(\tau)=0.002$ \cite{Watts18,Allys22,Valentino18}. At the tightest $\tau$ constraints from 21cm surveys \cite{Sailer22,Liu:2015txa,Qin22,Shmueli23}, the inclusion of LAEs will improve the measurements by 0.7$\sigma$ and 0.3$\sigma$ for Stage 4.5 and Stage V, respectively, allowing for the 5.5$\sigma$ detection of the normal mass hierarchy. 

The light relic forecasts on $N_{\rm eff}$ show that near future constraints from Stage 4.5 and Stage V will be comparable to or tighter than the Planck constraints. The assistance of multi-tracers are more significant with a BBN prior, with the total improvement at $38\%$. With a Planck prior, the help of both tracers decrease, marking a total improvement of 25\%. 
Finally, we see that the choice of Stage V design has minimal effect on the final constraints, providing errors tighter than that of Planck using only galaxy tracers and a BBN prior. 

\subsection{Primordial features}

Current constraints of inflation, primarily by CMB observations, are consistent with a simple, single scalar field slow-roll inflation. The (near) flatness of the inflationary potential leads to a simple explanation of the nearly scale-invariant primordial power-spectrum. When formulating a model from a more fundamental basis, however, one often finds deviation from said scale-invariance \cite{Sailer21,Slosar19c,Beutler19}. A canonical subclass of such models approximate deviation from scale-invariance over a narrow range of scales, using oscillations in $k$ or $\log{k}$. For this work, we model this as oscillations in the linear power-spectrum
\begin{equation}
    P_m(k)\rightarrow[1+A_{\rm lin}\sin{(\omega_{\rm lin}k+\phi_{\rm lin})}+A_{\rm log}\sin{(\omega_{\rm log}\log{k/k_*}+\phi_{\rm log})}]P_m(k)
\end{equation}
where $k_*=0.05\,h^{-1}\,\text{Mpc}$ is the pivot scale \cite{Beutler19,Ballardini16,Sailer21}. We fix $\phi_{\rm lin}$ and $\phi_{\rm log}$ to $\pi/2$ for all forecasts and characterize the change in constraining power as we vary $\omega_{\rm lin}$ and $\omega_{\rm log}$.

\begin{figure}
    \centering
    \includegraphics[width=\textwidth]{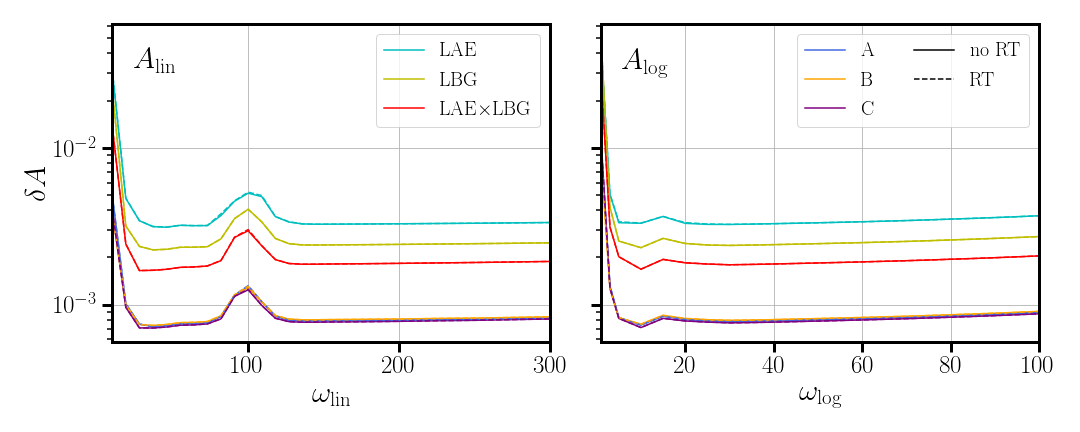}
    \caption{Forecast results for $A_{\rm lin}$ and $A_{\rm log}$ for a BBN-prior. `A', `B', and `C' each correspond to Plans A, B, and C of Stage V (Table \ref{tab:experiments}). The loss in $\delta A_{\rm lin}$ near $\omega_{\rm lin}=100$ with the BBN prior is due to the degeneracy with BAO \cite{Sailer21}, but this degeneracy is removed when using a Planck prior.}
    \label{fig:D2_SV_Alin_Alog}
\end{figure}

Our results, shown in Figures \ref{fig:D2_SV_Alin_Alog} indicate that use of multiple galaxy tracers benefits our sensitivity by 8\% for $A_{\rm lin}$ and 16\% for $A_{\rm log}$ with BBN priors, and less than 2\% for both with Planck priors. Furthermore, the inclusion of RT minimally affect the constraints, as increasing uncertainty in $\mu$ dependence does not directly alter our ability to detect $k$ dependence. We find that these constraints are robust against crude optimization strategies for Stage V. 

\subsection{Early dark energy}

While current measurements of cosmology are consistent with that of the cosmological constant $\Lambda$ as the source of dark energy, dynamical dark energy is an active area of research. Early dark energy (EDE) is among these models, with $\Omega_{\rm DE}$ rapidly dropping at high redshift. 
The specific model that we probe has the potential $V(\phi)\propto [1-\cos{(\phi/f})]^3$, where $\phi$ is a light scalar field \cite{Poulin19,Sailer21}. The field is parameterized by redshift $z_c$, where $f_{\rm EDE}\equiv\rho_{\rm EDE}/\rho_{\rm tot}$ is maximum, and the initial value of the field $\theta_i\equiv \phi_i/f$. This model, as well as other major EDE models, invokes a phase shift in BAO when $z_c\gtrsim z_*$, where $z_*\sim1100$ is recombination. Otherwise, when $z_c\lesssim z_*$, EDE damps the power-spectrum equally across scales for $k\gtrsim 0.01\,h\,\text{Mpc}^{-1}$ \cite{Sailer21}.
This can be understood by simply considering the derivation for sound horizon $r_s$ and angular diameter distance $D_A(z)$
\begin{equation}
    r_s =\int_{z_*}^\infty \frac{c_s(z)}{H(z)} dz \quad\quad
    D_A(z)=\int_0^z \frac{dz}{H(z)}
\end{equation}
When $z_c\gtrsim z_*$, the BAO scale is impacted by influence on $r_s$, while for $z_c\lesssim z_*$, there is no influence on either parameters (as long as $z<z_c$, which is always the case for realistic models and any quasi-near-time surveys).
The phase shifts on BAO, or more specifically modifications of the Hubble parameter measurements from the CMB, has the potential to relieve the Hubble tension. 
For the forecasts in this section, we fix $\theta_i=2.83$ and probe $1.5<\log_{10}(z_c)<6.5$ using \texttt{CLASS\_EDE} \cite{Hill20}.

\begin{figure}
    \centering
    \includegraphics[width=\textwidth]{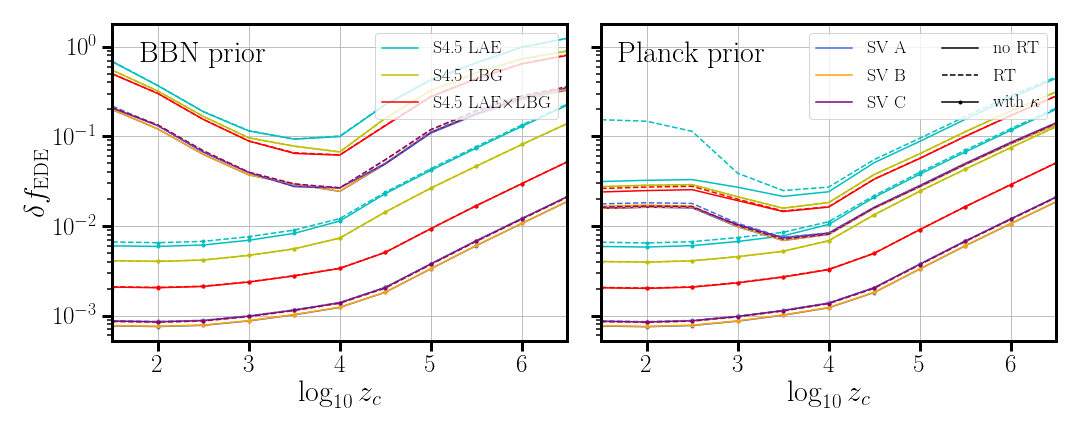}
    \caption{Forecast results for early dark energy. `S4.5' refers to Stage-4.5, whereas `SV A', `SV B', and `SV C' each correspond to Plans A, B, and C of Stage V (Table \ref{tab:experiments}). RT induces a loss of constraining power at with a Planck prior due to the additional amplitude degeneracy.}
    \label{fig:D2_SV_fEDE}
\end{figure}

The results shown in Figure \ref{fig:D2_SV_fEDE} indicate that multi-tracers play a significant role in EDE constraints. While the improvements through LAEs alone remain modest at $\lesssim10\%$ for both priors, lensing improves the constraints by $91\%$ with a BBN prior and $\sim64\%$ with a Planck prior, totalling together to 97\% and 85\%, respectively. 

We also observe that composite constraint is affected marginally by the presence of RT, while for LAE-only constraints this is not always true. 
Ref.~\cite{Sailer21} makes note that LSS-only constraints (without RT) are worsened at low $z_c$ ($z_c\lesssim z_*$) and attribute this to the degeneracy between the damping of power-spectra at $k\gtrsim 0.01\,h\,\text{Mpc}^{-1}$, $A_s$ and $b$. We recognize that for LAE-only constraints in the Planck prior case, RT significantly diminishes our constraining power further at this regime. This is entirely expected, as the variational $f$ at linear-theory level introduces an additional degree of freedom on the existing amplitude degeneracy. Notably, the inclusion of lensing nearly recovers our constraining power lost from RT. This again suggests that the inclusion of lensing aids the breaking of degeneracy introduced by RT. 

\section{Conclusions} \label{sec:conclusion}

Next generation experiments in large-scale structure aim to probe the high redshift universe to constrain the $\Lambda$CDM model and its extensions. Modelling of large-scale structure using perturbation theory introduces nuisance parameters (Table \ref{tab:fiducial}) such as bias $b_X$, counter-terms $\alpha_{2n}$, and stochastic terms $N_{2n}$ in order to make predictions in a controlled manner. A major limitation to cosmological constraints are the degeneracy that these parameters introduce in the data, as well as inherent degeneracies between cosmological parameters. 
Tracing multiple different tracers at the same redshift, called the ``multi-tracer'' technique, may allow us to break these degeneracies by using the different dependence of tracers to each parameter. Next generation facilities may choose to do this with Lyman alpha emitters (LAEs), Lyman break galaxies (LBGs), and CMB lensing $\kappa$. With different large-scale biases (Figures \ref{fig:LAEnb}, \ref{fig:LBGnb}, or the lack thereof), these tracers will be well-suited for multi-tracer cosmology.  In this work, we have compared forecasted cosmological constraints from single tracers, an ``incomplete'' multi-tracer technique (where one uses multiple galaxy tracers but only includes the two galaxy auto-spectra as signals neglecting the galaxy-galaxy cross-spectrum) and one which includes all available spectra. 

We start with a generic investigation of the mechanism of multi-tracer cosmology, using $\sigma_8$ forecasts for a set of simplified tracers. We find that degeneracy breaking indeed drives a majority of the improvement in constraining power, in agreement with recent studies \cite{Mergulhao22, Mergulhao23}. 

We then apply the multi-tracer technique for Fisher forecasts of next generation surveys, namely ``Stage 4.5'', inspired by DESI II, and ``Stage V'', using both BBN and Planck primary priors.
We find that the multi-tracer technique has varying effect on constraining power, as shown in Figure \ref{fig:summary}. 
In general, multi-tracer with LAEs on a LBG-only sample do not improve the constraints drastically with the improvements $\lesssim15\%$ for most parameters. Interestingly, when using LAEs as an addition onto a $\kappa\times$LBG sample, there are parameters that improve considerably, namely $h$ and $\omega_c$ with a Planck prior, at $\sim35\%$, and $f_{\rm EDE}$ at $\sim 60\%$. 
CMB lensing proves to be a better contribution to multi-tracers than LAEs in many cases, although the improvements cannot be compared on equal footing as they are adopted without any cost, while observing LAEs cost observation time that can otherwise be used for LBGs. With lensing, many of cosmological parameters being improved $\gtrsim20\%$, with a factor of $\gtrsim10$ improvement for $f_{\rm EDE}$ being the most drastic. 
Importantly, we find that these improvements aggregate well in general (e.g., for $\delta\sigma_8$ the $\sim10\%$ and $\sim20\%$ improvements from LAEs and lensing, respectively, lead to $\sim30\%$ improvements when combined), proving that we are not in a regime where there are enough tracers that additional tracers become insignificant.

Where they overlap our results agree well with the recent study of  ref.~\cite{Mergulhao23}, although the results are presented differently. Ref.~\cite{Mergulhao23} quotes $\sim50\%$ improvement in constraining power from multi-tracers with two galaxy samples, using $\sigma_{ST}/\sigma_{MT}-1$ as their metric.  We use $1-\sigma_{MT}/\sigma_{ST}$ which leads to smaller quoted improvements, e.g.\ a 50\% improvement in ref.~\cite{Mergulhao23} corresponds to a 33\% improvement in our metric, bridging a significant portion of the apparent gap between the final results. Remaining differences in results can be attributed to differences in the galaxy samples used, such as the difference between galaxy biases. A similar comment can be made for differences between this work and ref.~\cite{Zhao23}, which show improvements ranging from 5\% to 33\% from ELGs and LRGs. In both studies, there is a factor $\sim2$ difference between linear biases, whereas LAE and LBG biases differ only by 30\% at most (for Stage 4.5) in this work. 

We also perform a crude optimization of Stage V survey designs by changing luminosity limits for both LAEs and LBGs while keeping the total integration time fixed. We find that most Stage V constraints are robust to this optimization, leaving room for future optimization involving specific targets and observation techniques.  A detailed study of survey optimization, including galaxy mix as a function of redshift, time-to-redshift and redshift efficiency factors, is left for future work.

A possible source of concern for the use of LAEs as cosmological probes is the impact of radiative transfer (RT), specifically the absorption and scattering of Ly$\alpha$ emission lines by neutral hydrogen near the LAE. By assuming that RT effects are small, we attempt a preliminary investigation of the impact of RT by including its effects at the linear theory level, ignoring higher order contributions. Forecasts using this model show that $\sigma_8$ loses nearly half of its multi-tracer gains by LAEs, and LAE-only forecasts experience significant loss of constraining power for $\sigma_8$, $A_s$, neutrino mass, and early dark energy with Planck priors at low $z_c$ due to the additional amplitude degeneracy. On the contrary, however, we find that other forecasts are fairly unaffected, and all constraints experience little effect for the final, three-tracer forecast. Of course, this treatment of RT is merely preliminary and a more complete calculation will be attempted in the future.

\acknowledgments

We thank Shi-Fan Chen for discussion on counterterm treatment, and Arjun Dey, Simone Ferraro, Noah Sailer, and Ruiyang Zhao for useful discussions throughout the course of this work. 
H.E.~and M.W.~are supported by the DOE.
This research has made use of NASA's Astrophysics Data System and the arXiv preprint server.
This research is supported by the Director, Office of Science, Office of High Energy Physics of the U.S. Department of Energy under Contract No. DE-AC02-05CH11231, and by the National Energy Research Scientific Computing Center, a DOE Office of Science User Facility under the same contract.

\appendix

\section{Variance of measurements with correlation} \label{variance_calc}

In this section we walk through the calculation of the composite variance from two correlated measurements, leading to Eq.\ \ref{variance_linear}.
Suppose that two measurements $(x_1,x_2)$ are made drawing from a Gaussian distribution of mean $\mu$ and covariance $C$. The covariance matrix will be of the form 
\begin{equation}
    C=
    \begin{pmatrix}
        \sigma_1^2 & r\sigma_1 \sigma_2 \\
        r\sigma_1\sigma_2 & \sigma_2^2
    \end{pmatrix}
\end{equation}
where $r\in[-1,1]$ is the correlation coefficient. The probability of measuring $(x_1,x_2)$, then, is
\begin{equation}
    P(x_1,x_2)\propto \exp{(\Vec{x}-\Vec{\mu})^T C^{-1}(\Vec{x}-\Vec{\mu})/2}
\end{equation}
where $\Vec{x}=(x_1,x_2)$ and $\Vec{\mu}=(\mu,\mu)$. Maximizing the likelihood of observation provides the best estimate of $\mu$, with variance $\sigma^2$
\begin{equation}
    \mu = \frac{\Vec{1}^T C^{-1}\Vec{x}}{\Vec{1}^T C^{-1}\Vec{1}},
    \quad\quad
    \sigma^2 = \frac{(1-r^2)\sigma_1^2 \sigma_2^2}{\sigma_1^2-2r\sigma_1\sigma_2+\sigma_2^2}
\end{equation}
where $\Vec{1}=(1,1)$.
Taking some limits with these expressions reveal important insights. In the fully uncorrelated case $r=0$, one recovers the standard weighted average $\mu=(x_1/\sigma_1^2+x_2/\sigma_2^2)(1/\sigma_1^2+1/\sigma_2^2)^{-1}$, $\sigma^2 = (1/\sigma_1^2+1/\sigma_2^2)^{-1}$. When two measurements are redundant, with $\sigma_1=\sigma_2$ and $r\to1$, one obtains $\sigma^2=\sigma_1^2$, as expected. 
The best case is where there is near full correlation $\abs{r}\to$, with non-redundant information $\sigma_1\ne\sigma_2$. Then, one obtains $\sigma^2\to 0$. 

\section{Conventions of limiting luminosity in literature}
\label{luminosity}

Literature from the astronomy community often use two different conventions to make note of their luminosity function integration range. In this section we show the conversion between the two conventions.

The first approach is to simply use the lower limit of the integral, e.g.
\begin{equation}
    \Bar{n}=\int_{L_{\rm lim}}^\infty \Phi(L') dL'
\end{equation}
Another is to use the median, e.g. $L_{\rm med}$, such that
\begin{equation}
    \int_{L_{\rm med}}^\infty \Phi(L') dL'=\frac{1}{2}\int_{L_{\rm lim}}^\infty \Phi(L') dL'
\end{equation}
Note that in some authors place an explicit, finite upper limit on the integrals depending on the context.
These definitions are extended for limiting line fluxes or magnitudes analogously. It is clear that with the luminosity function (LF) parameters, one can convert one to the other easily. For the regimes of our interest, the differences are typically $\sim0.1-0.2$ in log-space. 

\section{Counterterms}
\label{sec:counterterms}

The theory of Lagrangian EFT has been developed in refs.~\cite{Zel70,Buc89,Mou91,Hivon95,TayHam96,ZheFri14,Mat15,McQWhi16,VlaCasWhi16,Mat08a,Mat08b,CLPT,White14,PorSenZal14,VlaWhiAvi15,Vlah16,VlaWhi19}.  However in those works the cross-spectra between different tracers was not highlighted.  In this appendix we briefly recap the key points of Lagrangian EFT, with a focus on the counterterms for the cross-spectra which have not appeared explicitly before. This will also allow us to make contact with the Eulerian approach of ref.~\cite{Mergulhao23}, which has the same counterterm structure.

We will perform calculations at both the density field-level and the power-spectrum-level. The latter is a more precise follow-through of the calculation, as the expectation values are taken explicitly, but we will see that the former will yield the same result with a simpler interpretation. 

We will start with the field-level interpretation. 
In Lagrangian EFT, the density field of a tracer is given by 
\begin{equation}
    \delta_a + (2\pi)^3\delta_D(\textbf{k})= \int_{\textbf{q}} F[\delta(q)] \,e^{i\textbf{k}\cdot(\textbf{q}+\Psi(\textbf{q}))}
    \label{density}
\end{equation}
where $\Psi$ is the displacement field and the functional $F$ describes the tracer's dependence on underlying fields using bias. For this section, we limit ourselves to linear order, $F=1+b_1\delta$, with $b_1$ the Lagrangian bias. Including higher-order terms ($\delta^2$, $s^2$, etc.) will not add to this section as their contributions to the counterterm will exceed one-loop order. For similar reasons we will limit ourselves to the linear solution of $\Psi$. 

To obtain the counterterms, we begin by dividing the displacements into long-wavelength components and a short-wavelength component. The former are treated perturbatively, while the latter can only be constrained by symmetry arguments and are included as a derivative expansion. 
Some of these terms correlate with the long-wavelength contributions while some do not. We refer to the former as counterterms and the latter as stochastic terms. 
These terms allow for the final results to be independent of the cutoff scale $\Lambda$ used in defining the perturbative terms. 

To linear order the (long-wavelength) displacement field $\Psi^L$ is simply $(i\vec{k}/k^2)\,\delta_{\rm lin}(k)$ \cite{Zel70}. In redshift space we apply an additional displacement along the line of sight \cite{Kai87}. Since $\dot{\Psi}^L=f\Psi^L$, in Hubble units, for the linear displacement the transition to redshift space is accomplished by multiplying the line-of-sight component of $\Psi^L$ by $1+f$. 
Then, in the absence of counterterms (i.e.\ absence of small-scale contributions), we obtain the Kaiser power spectrum by expanding the exponent $e^{i\textbf{k}\cdot\Psi}$ as 
\begin{equation}
    e^{i\textbf{k}\cdot\Psi(s)} \simeq 1+i\textbf{k}\cdot\Psi(s)
    = 1+i\textbf{k}\cdot [\Psi(q)+f\hat{n} (\hat{n}\cdot\Psi(q))]
\end{equation}
where $q$ and $s$ are the real-space and redshift-space coordinates, respectively. Substituting this to Eq.~\ref{density} and using $\mathbf{k}\cdot\hat{n}=k\mu$ yields $\delta_a =([1+b_1^{(a)}]+f\mu^2)\delta_{\rm lin}$, where $1+b_1$ is equivalent to linear bias in Eulerian PT.

Now let us divide our displacement field into $\Psi = \Psi^L + c_0(t) \nabla\delta_{\rm lin} + \epsilon_i + \cdots$. The term $c_0\nabla\delta_{\rm lin}$ is the first-order counterterm (i.e.~due to small scale dynamics) contribution to $\Psi$ allowed by symmetry, whereas $\epsilon$ is the stochastic noise. As the time-dependence of $c_0$ and $\epsilon$ are unknown, we gain additional degrees of freedom for $\dot{\Psi}$. Let us define this as $c_2$ and $\dot\epsilon$ such that
\begin{align}
    \label{expandPsi}
    \Psi_i&=\Psi_i^L+c_0\nabla\delta_{\rm lin}+\epsilon_i\\
    \label{expandPsi2}\dot\Psi_i&=\dot\Psi_i^L+c_2\nabla\delta_{\rm lin}+\dot\epsilon_i
\end{align}
Performing a direct expansion for small-scale contributions as we did for linear theory calculations, we find
\begin{align}
    \delta_a&\supset \int_{\textbf{q}}e^{i\textbf{k}\cdot\textbf{q}} \sum_n\frac{1}{n!} [ik_i (\Psi_i+\dot\Psi_{\hat{n},i})]^n \\
    &\supset\int_{\textbf{q}}e^{i\textbf{k}\cdot\textbf{q}}  \left(-c_{a,0}-c_{a,2}\mu^2+\frac{1}{2}(1+f\mu^2)((\hat{k}_i\epsilon_i)^2+(\hat{n}_i\dot\epsilon_i)^2 \mu^2)\right)k^2\delta_{\rm lin}
\end{align}
where we use $\dot{\Psi}^L_{\hat{n},i}=\hat{n}_i\hat{n}_j \dot{\Psi}^L_j$ for shorthand and the subscripts $a$ on coefficients allow for tracer-dependent small-scale dynamics (whereas stochastic contributions $\epsilon$ do not have tracer-dependence). Note that we have neglected terms that go as even powers of $\delta_{\rm lin}$ or odd powers of $\epsilon,\,\dot\epsilon$ as they vanish when taking expectation values at the power-spectrum level.

To lowest order the counterterms for the autospectrum of tracer $a$ are thus $k^2\,P_{\rm lin}$ times
\begin{equation}
    2\left( [1+b_1^{(a)}]+f\mu^2 \right)\left(-\tilde{c}_{a,0}+ -\tilde{c}_{a,2}\mu^2 - \tilde{c}_4\mu^4 \right)
\end{equation}
while those for the cross-spectrum between tracers $a$ and $b$ are $k^2\,P_{\rm lin}$ times
\begin{equation}
    \left( [1+b_1^{(a)}]+f\mu^2 \right)\left( -\tilde{c}_{b,0}+ -\tilde{c}_{b,2}\mu^2 - \tilde{c}_4\mu^4\right)+
    \left( [1+b_1^{(b)}]+f\mu^2 \right)\left( -\tilde{c}_{a,0}+ -\tilde{c}_{a,2}\mu^2 - \tilde{c}_4\mu^4\right)
\end{equation}
where we have redefined coefficients for simplicity. 
The cross-spectra thus introduce no new free parameters beyond those already needed for the counterterms of the auto-spectra, and this increases the utility of the cross-spectrum in breaking degeneracies. Note that although $\tilde{c}_4$ is a shared contribution across tracers, in this work we will allow them to be tracer-dependent, in order to be conservative. We expect that this will change the results minimally. 

Now, we will perform the calculation at the power-spectrum-level to see that our results above are indeed correct. To do this, we will follow the moment expansion approach \cite{VlaWhi19,SelMcD11,Vla13}. In this approach, one uses the fact that the redshift space power-spectrum is a specific case of the moment generating function and expands it such that 
\begin{align}
    \frac{k^3}{2\pi^2}P_s(\textbf{k})=\Tilde{M}(\textbf{J}=\textbf{k},\textbf{k})&=\frac{k^3}{2\pi^2}\int d^3r e^{i\textbf{k}\cdot\textbf{r}} \expval{(1+\delta_g(\textbf{x}_1))(1+\delta_g(\textbf{x}_2))e^{i\textbf{k}\cdot\Delta\textbf{u}}}_{\mathbf{x}_1-\mathbf{x}_2=\mathbf{r}}\\
    &= \frac{k^3}{2\pi^2}\sum_{n=0}^\infty \frac{i^n}{n!}k_{i_1}\dots k_{i_n}\tilde\Xi_{i_1\dots i_n}^{(n)}(\textbf{k})
\end{align}
where $\mathbf{u}=\hat{n}(\hat{n}\cdot\mathbf{v})/\mathcal{H}$ is the LOS velocity, $\Delta\mathbf{u}=\mathbf{u}(\mathbf{x}_1)-\mathbf{u}(\mathbf{x}_2)$ their difference, and $\tilde\Xi^{(n)}$ are Fourier transforms of the velocity moments
\begin{equation}
    \Xi_{i_1\dots i_n}^{(n)} = \expval{(1+\delta_1)(1+\delta_2) \Delta\textbf{u}_{i_1}\dots\Delta\textbf{u}_{i_n}}
    \quad .
\end{equation}
Notice that here, the first moment is simply the real-space power-spectrum and the higher moment contributions provide information about redshift space distortions (RSD). As RSD depend only on the line-of-sight component of velocities, the only non-vanishing contributions to $P_s$ come from the LOS component of $\tilde\Xi$. With this in mind, we rewrite the equation above as
\begin{equation}
    P_s(\textbf{k})=\sum_{n=0}^\infty \frac{i^n}{n!}(k\mu)^n \sum_{\ell=0}^n \tilde\Xi_\ell^{(n)}\mathcal{L}_\ell(\mu) 
\end{equation}
where $\tilde\Xi_\ell^{(n)}$ are Legendre moments of the velocity moments.

Formally, counterterm contributions arise from UV-divergences that arise from contractions of operators at small distances. As our theory knows only of large-scale dynamics, the only constraints we have on these contributions are from symmetry. Those that appear in the final expression are
\begin{align}
    \expval{\Psi_i \Psi_j} &\supset \beta_1  \delta_{ij}+\beta_2  \delta_{ij} \delta + \beta_3 \hat{k}_i\hat{k}_j \delta \\
    \expval{\delta \Psi_i} &\supset \beta_4 ik_i \delta 
\end{align}
and similarly for contact terms with derivatives, e.g.~$\Psi_i\dot\Psi_j\supset\dot{\beta}_1\delta_{ij}+ \dot{\beta}_2\delta_{ij}\delta+\dot{\beta}_3\hat{k}_i\hat{k}_j \delta$. Note that the coefficients $\beta$ will depend on the fields in contact, e.g.~$\Psi_i^a\Psi_j^b\supset\beta_1^{ab}\delta_{ij}$ and $\beta_1^{aa}\ne\beta_1^{ab}$. However, these will not yield new degrees of freedom for the cross-spectra since the underlying mechanism is tracer dependent, as  described in Eq. \ref{expandPsi}-\ref{expandPsi2}.

Using the above contributions, one can compute the contributions to each velocity moment multipole. For example, we have 
\begin{align}
    \dddot{W}_{ijk}&=\expval{\dot\Delta_i\dot\Delta_j\dot\Delta_k}\\
    &=\underbrace{\expval{\dot\Psi_i(q_1)\dot\Psi_j(q_1)\dot\Psi_k(q_1)-\left(q_1\to q_2\right)}}_{\propto \expval{\delta}=0\,\text{for c.t.}} + \expval{\dot\Psi_{\{i}(q_1)\dot\Psi_j(q_2)\dot\Psi_{k\}}(q_2)-(q_1\leftrightarrow q_2)}\\
    &\supset \expval{\frac{2i k_{\{i}}{k^2}\delta_1\left(\ddot{\beta}_1 \delta_{jk\}}+\ddot{\beta}_2 \delta_{jk\}} \delta_2 + \ddot{\beta}_3 \hat{k}_j\hat{k}_{k\}} \delta_2\right)} \\
    &=\frac{2i}{k}P_Lf\left(\ddot{\beta}_2 \hat{k}_{\{i}\delta_{jk\}}+3\ddot{\beta}_3\hat{k}_{i}\hat{k}_{j}\hat{k}_{k} \right)
\end{align}
Using the velocity moment expressions \cite{Chen20a}, we see that this contributes to the velocity moment multipoles
\begin{equation}
    -i\tilde\Xi^{(3)}_{1,ct} = \frac{6}{k}P_L f\left(\ddot{\beta}_2 +\frac{3}{5}\ddot{\beta}_3\right)
    \quad , \quad
    -i\tilde\Xi^{(3)}_{3,ct} = \frac{12}{5k}P_L f\ddot{\beta}_3
\end{equation}
which in turn contributes
\begin{equation}
    P_{s,ct}\supset k^2P_L(\Ddot{\beta}_2f\mu^4+\ddot{\beta}_3f\mu^6)
\end{equation}
Note that for $n=3$ we use the Legendre basis instead of the polynomial basis shown in the Appendix of \cite{Chen20a}.
Performing a similar computation for all terms, one obtains
\begin{equation}
    P_{s}\supset k^2P_L (\tilde\beta_1 + \beta_2+\beta_3-2b \beta_4 - 2c_0 + (\dot{\tilde\beta}_1 + 2\dot{\beta}_2+2\dot{\beta}_3-2b\dot{\beta}_4+\ddot{\beta}_2 - 2c_2)\mu^2+(\ddot{\tilde\beta}_1 + \ddot{\beta}_3)\mu^4)(1+b+f\mu^2)
\end{equation}
in agreement with the field-level calculation above, where $\beta_1$'s have been redefined for brevity\footnote{Note that $\ddot{\tilde\beta}_1$ is determined by $b$, $\tilde\beta_1$, and $\dot{\tilde\beta}_1$, but this will not reduce the degrees of freedom due to degeneracy with $\ddot\beta_3$} and terms with $c_0$ and $c_2$ (as defined in Eq. \ref{expandPsi}-\ref{expandPsi2}) come from non-contact counterterms. 

\bibliographystyle{JHEP}
\bibliography{main}

\end{document}